\documentclass[11pt,english]{article}

\usepackage[utf8]{inputenc} 

\usepackage{dsfont}

\usepackage[left=1in,right=1in,top=1in,bottom=1in]{geometry}
\usepackage{authblk}
\usepackage{enumitem}
\usepackage{ulem} 
\usepackage{empheq}
\usepackage[none]{hyphenat}

\usepackage{graphicx}
\usepackage{subfigure}
\usepackage{float}
\usepackage{wrapfig} 

\usepackage{xcolor}
\usepackage{color}
\definecolor{darkgreen}{rgb}{0,0.5,0}
\definecolor{darkblue}{rgb}{0,0,0.6}
\definecolor{purple}{rgb}{0.4,.2,0.7}
\usepackage{colortbl}      

\usepackage{amsmath}
\usepackage{amssymb}

\usepackage{tensor} 
\usepackage{slashed}
\usepackage{mathtools}
\usepackage{leftidx}
\usepackage{esint}
\usepackage{amsfonts,amsthm,bm}

\usepackage{physics}
\usepackage{feynmf}
\usepackage{braket}
\usepackage{ytableau}

\usepackage{prettyref}
\usepackage[colorlinks=true,citecolor=darkgreen,linkcolor=purple,urlcolor=purple]{hyperref}

\usepackage{tikz}

\usepackage[numbers,sort&compress]{natbib}

\newcommand{\nn}{\nonumber}

\numberwithin{equation}{section}
\numberwithin{figure}{section}
\numberwithin{table}{section}

\begin{document}

\emergencystretch 3em

\title{\LARGE\textsc{Gravitons on Nariai Edges}}

\author[a]{\vskip1cm \normalsize Y.T.\ Albert Law}
\affil[a]{ \it \normalsize Leinweber Institute for Theoretical Physics at Stanford, 382 Via Pueblo, Stanford, CA 94305, USA}
\author[b]{\normalsize Varun Lochab}
\affil[b]{ \it \normalsize Center for Theoretical Physics, Columbia University, New York, NY 10027, USA}
\date{}
\maketitle

\begin{center}
	\vskip-10mm
	{ \href{mailto:ytalaw@stanford.edu}{ytalaw@stanford.edu}, \; \href{mailto:vv2338@columbia.edu}{vv2338@columbia.edu}}
\end{center}

\vskip10mm

\thispagestyle{empty}

\begin{abstract}
We show that, for any $d\geq 3$, the one-loop graviton path integral on $S^2\times S^{d-1}$ factorizes into bulk and edge parts. The bulk equals the thermal partition function of an ideal graviton gas in the Lorentzian Nariai geometry. The edge factor is the inverse of the path integral over two identical copies, each containing one shift-symmetric vector and three shift-symmetric scalars on $S^{d-1}$. Unlike the round $S^{d+1}$ case, all scalars are massless, indicating that graviton edge partition functions probe beyond the horizon’s intrinsic geometry---in contrast to $p$-form gauge theories. In the course of this work, we obtain a compact formula for the one-loop Euclidean graviton path integral on any $\Lambda >0$ Einstein manifold.

\end{abstract}


\newpage

\tableofcontents



\section{Introduction}

In quantum gravity with a positive cosmological constant, a natural field-redefinition and diffeomorphism-invariant object is the Euclidean gravitational path integral
\begin{align}\label{introeq:Z}
\mathcal{Z} = \int \mathcal{D}g \, e^{-S[g]} \;.
\end{align}
Inspired by studies of black holes and the AdS/CFT correspondence, one might hope that understanding how to define or compute \eqref{introeq:Z} could shed light on the underlying microscopic theory. Unlike those cases, however, the absence of a natural asymptotic boundary on which the microscopic theory ``lives” makes the path toward interpreting \eqref{introeq:Z} far from obvious.

In any dimension $D = d+1 \geq 3$, the leading saddle point of \eqref{introeq:Z} is the round $S^{d+1}$. The 1-loop corrections from free field fluctuations around this geometry were analyzed in \cite{Anninos:2020hfj}. Remarkably, the 1-loop $S^{d+1}$ partition functions for any free field were found to exhibit a universal structure:
\begin{align}\label{introeq:PIsplit}
	Z^\text{1-loop}_\text{PI}  \left[S^{d+1} \right]=Z_{\rm bulk} \left(\beta=\beta_\text{dS}\right) Z_{\rm edge}\;. 
\end{align}
Specializing to gravitons, the first factor is given by
\begin{align}\label{introeq:Zbulk}
	\log Z_\text{bulk} (\beta)\equiv\int_0^\infty \frac{dt}{2t} \frac{1+e^{-2\pi t/\beta}}{1-e^{-2\pi t/\beta}}\,\chi(t) \; , \quad \chi(t)	
	=  \sum_{I=0}^2 \sum_{n=0}^\infty \sum_{l=2}^\infty D^d_{l,I} \left(q^{I+2n+l} +q^{d-I+2n+l}\right) \;,
\end{align}
where $q\equiv e^{-|t|/\ell_\text{dS}} $. The quantity $\chi(t)$ is the Harish-Chandra character of the de Sitter (dS) boost generator for the massless spin-2 representation of $SO(1,d+1)$ \cite{10.3792/pja/1195523378,10.3792/pja/1195523460,10.2307/1992907,bams/1183525024,10.3792/pja/1195522333}. The exponents in the sum correspond to the physical quasinormal mode (QNM) frequencies (times $i$) for gravitons on a $dS_{d+1}$ static patch, with degeneracies $D^d_{l,I}$ defined in appendix \ref{sec:harmonics}. Alternatively, $\chi(t)$ can be understood as a spatially integrated Green function \cite{Grewal:2024emf,spin}. One can view \eqref{introeq:Zbulk} as the thermal canonical partition function of an ideal-gas defined on the continuous normal mode spectrum, with spectral measure\footnote{$\tilde\rho(\omega)$ can be understood as a relative or renormalized spectral density, defined in terms of scattering phases associated with the reduced scattering problems descending from the free field equations \cite{Law:2022zdq}. A brief review in the dS context can be found in \cite{Law:2023ohq}.} 
\begin{align}
    \tilde\rho(\omega)=\int_{-\infty}^\infty \frac{dt}{2\pi} e^{-i\omega t}\chi(t)\;.
\end{align}
In \eqref{introeq:PIsplit}, the inverse dS temperature is $\beta_\text{dS} = 2\pi \ell_\text{dS}$, where $\ell_\text{dS} \equiv \sqrt{ \frac{d(d-1)}{2\Lambda} }$ is the dS radius.

On the other hand, the second, ``edge'', factor in \eqref{introeq:PIsplit} takes the form of a path integral on $S^{d-1}$. Recently, using the branching rule $SO(d+2)\to U(1)\times SO(d)$, \cite{Law:2025ktz} refined the original formula of \cite{Anninos:2020hfj}, expressing $Z_{\rm edge}$ in terms of functional determinants on $S^{d-1}$:\footnote{The primes indicate exclusion of zero modes from the determinants. The subscript $-1$ on the vector determinant refers to a continuation of the $SO(d)$ angular momentum index $l$ from 1 to $-1$, as explained in \cite{Law:2025ktz}. The parameter $M$ has dimensions of mass and renders the expression dimensionless. The proportionality constant contains additional non-local contributions to $Z_{\rm edge}$.}
\begin{align}\label{introeq:Zedgegravresult}
    Z_{\rm edge }
    \propto \det\nolimits'_{-1} \left| \frac{-\nabla_1^2-\frac{d-2}{\ell^2_\text{dS}}}{M^2} \right|^{\frac12}  \det\nolimits' \left| \frac{-\nabla_0^2-\frac{d-1}{\ell^2_\text{dS}}}{M^2} \right| \det\nolimits'  \left(\frac{-\nabla_0^2}{M^2} \right)^{\frac12} \;.
\end{align}
Here, $-\nabla_0^2$ and $-\nabla_1^2$ are Laplacians acting on scalars and transverse vectors on $S^{d-1}$, respectively. One sees that this expression receives ghost-like contributions from one tachyonic vector, two tachyonic scalars, and one massless scalar. A possible interpretation in terms of an $S^{d-1}$ brane embedded in $S^{d+1}$ was proposed in \cite{Law:2025ktz}. While \eqref{introeq:Zedgegravresult} likely admits a Lorentzian interpretation in terms of gravitational edge modes \cite{Donnelly:2016auv,Geiller:2017xad,Speranza:2017gxd,Geiller:2017whh,Freidel:2019ees,Takayanagi:2019tvn,Freidel:2020xyx,Freidel:2020svx,Freidel:2020ayo,Donnelly:2020xgu,Ciambelli:2021vnn,Carrozza:2021gju,Ciambelli:2021nmv,Carrozza:2022xut,Ciambelli:2022cfr,Mertens:2022ujr,Wong:2022eiu,Donnelly:2022kfs,Lee:2024etc,blommaert2025gravitons,Fliss:2024don}— analogous to those in Maxwell \cite{Kabat:1995eq,Donnelly:2011hn, Donnelly:2012st, Eling:2013aqa, Radicevic:2014kqa, Donnelly:2014gva, Donnelly:2014fua, Huang:2014pfa, Ghosh:2015iwa, Hung:2015fla, Aoki:2015bsa, Donnelly:2015hxa, Radicevic:2015sza, Pretko:2015zva, Soni:2015yga,Zuo:2016knh, Soni:2016ogt, Delcamp:2016eya, Agarwal:2016cir, Blommaert:2018rsf, Blommaert:2018oue, Freidel:2018fsk,Ball:2024hqe,Araujo-Regado:2024dpr} and $U(1)$ $p$-form gauge theories \cite{Dowker:2017flz, Moitra:2018lxn,David:2021wrw, Mukherjee:2023ihb, Dowker:2024cwy,Ball:2024xhf}—further work is needed to fully establish this connection.

In this work, we extend the story for any $d \geq 3$ to another saddle of \eqref{introeq:Z}, namely the product geometry $S^2 \times S^{d-1}$. While the round $S^{d+1}$ admits a Lorentzian continuation as a static patch in $dS_{d+1}$, $S^2 \times S^{d-1}$ continues instead to a static Nariai black hole, thereby offering a setting in which to contrast dS and black hole horizons for generic dimensions.\footnote{When $d+1=3$, 1-loop graviton partition functions and their bulk-edge split have been worked out for rotating BTZ black holes in \cite{Kapec:2024zdj}.} Moreover, it represents the simplest case featuring two disconnected horizons. A concise summary of these results, together with broader perspectives, appears in \cite{Law:2026tuk}.

In section \ref{sec:Lorentzian}, we study free gravitons on the Lorentzian Nariai geometry. Working in the transverse traceless (TT) gauge, we explicitly solve the graviton equations of motion and obtain the normal mode spectrum in section \ref{sec:normalmode}. In section \ref{sec:QNMZbulk}, we follow the approach of \cite{Law:2022zdq,Grewal:2022hlo} to define a spectral density on the continuous graviton normal mode spectrum, which in turn allows us to construct an ideal-gas thermal partition function \eqref{eq:bulkcanfn} at arbitrary inverse temperature $\beta$, in direct analogy with \eqref{introeq:Zbulk}.

In section \ref{sec:EinsPI}, we briefly digress to present a detailed analysis of the 1-loop path integral for pure Einstein gravity on an arbitrary closed Einstein manifold $\mathcal{M}$, following the approach of \cite{Law:2020cpj}. While not strictly necessary for our main purpose, this general perspective clarifies the structure of the 1-loop gravitational path integral and will serve as a foundation for later sections. We derive a general expression in terms of determinants of Laplace-type operators on $\mathcal{M}$, with careful treatment of zero and negative modes. The overall phase of the path integral has recently been discussed in \cite{Shi:2025amq,Ivo:2025yek}. Our final result is summarized in \eqref{eq:ZPIgenEinstein}. 

After setting the stage with this general analysis, we specialize to the case of $S^2 \times S^{d-1}$ and study the 1-loop graviton path integral in section \ref{sec:PINariai}. Using the Laplacian spectra on $S^2\times S^{d-1}$ derived in appendix \ref{sec:spec}, we obtain the bulk-edge factorization of $Z^\text{1-loop}_\text{PI}  \left[S^2 \times S^{d-1} \right]$, analogous to \eqref{introeq:PIsplit}, with $Z_\text{edge}$ given explicitly in \eqref{eq:Zedgedet}. The physical interpretation of this result is discussed in section \ref{sec:discussion}.

Appendix \ref{sec:STSH} reviews essential facts about spherical harmonics on $S^{d-1}$ and sets up our notational conventions, followed by a spectral analysis of Laplace-type operators on $S^2\times S^{d-1}$.


\section{Gravitons on the Lorentzian Nariai geometry}\label{sec:Lorentzian}

The Nariai geometry arises as the limit of a Schwarzschild–de Sitter black hole in which the cosmological and black-hole horizons coincide. Zooming into the region between the two horizons, the geometry becomes the direct product $dS_2\times S^{d-1}$ \cite{1950SRToh..34..160N}
\begin{align}\label{eq:Nariaimetric}
ds^2=-\left(1-\frac{\rho^2}{\ell_N^2}\right) dt^2 +\frac{d\rho^2}{1-\frac{\rho^2}{\ell_N^2}}+r_N^2 \, d\Omega_{d-1}^2\;, \qquad -\ell_N < \rho <\ell_N \; .
\end{align}
Here $d\Omega_{d-1}^2$ is the metric on a unit round $S^{d-1}$, and
\begin{align}\label{eq:lNcc}
    \ell_N \equiv \sqrt{\frac{d-1}{2\Lambda}}  \equiv \frac{r_N}{\sqrt{d-2}}\; ,
\end{align}
where $\ell_N$ is the dS length of the $dS_2$ factor, and $r_N$ is the radius of the transverse sphere $S^{d-1}$. In \eqref{eq:Nariaimetric}, the coordinate $t$ is the proper time of an observer located at $\rho=0$, who is surrounded by the cosmological and black hole horizons at $\rho=\pm \ell_N$ and who experiences the Hawking temperature
\begin{align}\label{eq:Nariaitemp}
T_N=\frac{1}{2\pi \ell_N} 
\end{align}
in the global vacuum state. Note that \eqref{eq:Nariaitemp} is higher than the pure dS temperature $T_\text{dS}\equiv \frac{1}{2\pi \ell_\text{dS}}$.

The only non-vanishing components of the Riemann tensor for \eqref{eq:Nariaimetric} are those with all indices in the $dS_2$ factor or all in the $S^{d-1}$ factor:
\begin{align}\label{eq:RiemNariaiL}
		R_{abcd} = \frac{g_{ac}g_{bd}-g_{bc}g_{ad}}{\ell_N^2} \;, \qquad 
		R_{ijkl}  = \frac{g_{ik}g_{jl}-g_{jk}g_{il}}{r_N^2} \;.
\end{align}
Here $a,b,c,d$ are $dS_2$ indices and $i,j,k,l$ are $S^{d-1}$ indices. The Ricci scalar is
\begin{align}\label{eq:RicciscalarNariaiL}
	R = \frac{2}{\ell^2_N} + \frac{(d-1)(d-2)}{r^2_N} =  \frac{(d+1)(d-2)}{r^2_N} = \frac{d+1}{\ell_N^2}\;. 
\end{align}

\subsection{Physical graviton normal modes}\label{sec:normalmode}

We are interested in gravitational waves propagating on the background Nariai geometry \eqref{eq:Nariaimetric}. These are captured by the linearized Einstein equation in the transverse traceless (TT) gauge\footnote{There are non-radiative solutions of the linearized Einstein equation that cannot be brought to the TT gauge, such as a static, spherically symmetric perturbation shifting the black-hole mass away from its Nariai value \cite{Ginsparg:1982rs,Bousso:1997wi}.}
\begin{align}\label{eq:EOMTT}
    - \nabla^2  h_{\mu \nu}
 -2 R_{\mu\alpha\nu\rho} h^{\alpha\rho} =0 \;, \qquad \nabla^\lambda h_{\lambda \mu} =0 = h\indices{^\lambda_\lambda} \;. 
\end{align}
We work in the TT gauge because it makes comparison with the Euclidean modes on $S^2\times S^{d-1}$ obtained in appendix \ref{sec:spec} transparent. 

The system \eqref{eq:EOMTT} is invariant under the residual gauge transformation
\begin{align}
    h_{\mu\nu} \to h_{\mu\nu} +\nabla_\mu \xi_\nu +\nabla_\nu \xi_\mu \;, 
\end{align}
where the gauge parameter satisfies the Proca equation of motion with Lagrangian mass $m^2=-\frac{2}{\ell^2_N}$
\begin{align}\label{eq:ghostsystem}
    \left(-\nabla^2-\frac{R}{d+1}\right) \xi_\mu = 0 \;, \qquad \nabla^\lambda \xi_\lambda =0 \;.
\end{align}
In what follows, we determine the spectrum of physical normal modes—those with time dependence $\propto e^{-i\omega t}$ and $\omega >0$—from which one can build normalizable wave packets. 

Since the geometry \eqref{eq:Nariaimetric} factorizes as $dS_2\times S^{d-1}$, expanding in spherical harmonics on $S^{d-1}$ (reviewed in appendix \ref{sec:STSH}) yields an infinite Kaluza–Klein tower of equations for (massive) scalars and vectors on $dS_2$, with masses labeled by the $SO(d)$ angular momentum $l$. Consequently, the problem reduces to finding normal modes of (massive) scalars and vectors on $dS_2$ static patch,
\begin{align}\label{eq:dS2}
ds^2=-\left(1-\frac{\rho^2}{\ell_N^2}\right)d t^2 +\frac{d\rho^2}{1-\frac{\rho^2}{\ell_N^2}}\;, \qquad -\ell_N < \rho <\ell_N \; .
\end{align}

\paragraph{Scalars on $dS_2$}

For a scalar of mass $m^2 \ell^2_N\equiv \Delta (1-\Delta) $ on \eqref{eq:dS2}, the Klein-Gordon equation
\begin{align}\label{eq:KGdS2}
    \left(-\nabla^2_{dS_2} + \frac{\Delta \bar \Delta}{\ell^2_N}  \right)\phi =0 \;, \qquad \bar \Delta \equiv 1-\Delta \;,
\end{align}
reduces to a one-dimensional  Schr\"{o}dinger problem with a P\"{o}schl-Teller-like potential. It admits two linearly independent normal mode solutions that are smooth everywhere on \eqref{eq:dS2}:\footnote{A Legendre-function basis is
\begin{align}
    R^1_{\omega \Delta}(t,\rho) &= e^{-i\omega t} P^{i\omega}_{-\Delta}\left(\frac{\rho}{\ell_N} \right)\qquad \text{and} \qquad R^2_{\omega \Delta}(t,\rho) = e^{-i\omega t} Q^{i\omega}_{-\Delta}\left(\frac{\rho}{\ell_N} \right) \;,
\end{align}
which manifests the connection with spherical harmonics on $S^2$. Under the analytic continuation $ t \to -i \tau$,  $\omega \to -im$ and $\Delta \to -L $, $R^1_{\omega \Delta}$ becomes the regular harmonic $Y_{Lm}\propto e^{im\tau}P^m_L \left(\frac{\rho}{\ell_N} \right)$, whereas $R^2_{\omega \Delta}$ maps to the singular partner $ e^{im\tau}Q^m_L \left(\frac{\rho}{\ell_N} \right)$, which diverges at $\rho=\pm \ell_N$.}
\begin{align}\label{eq:ds2solutions}
    f^\text{even}_{\omega \Delta}(t,\rho) &= e^{-i\omega t} \left(1 - \frac{\rho^2}{\ell_N^2}\right)^{-\frac{i\omega \ell_N}{2}}
\,_2F_1\left(\frac{\Delta - i\omega\ell_N}{2}, \frac{\bar{\Delta} - i\omega\ell_N}{2}; \frac12; 
\frac{\rho^2}{\ell_N^2}\right)\nn\\
    f^\text{odd}_{\omega \Delta}(t,\rho) &= e^{-i\omega t} \left(\frac{\rho}{\ell_N}\right)\,
\left(1 - \frac{\rho^2}{\ell_N^2}\right)^{-\frac{i\omega \ell_N}{2}}
\,_2F_1\left(\frac{1+\Delta - i\omega\ell_N}{2}, \frac{1+\bar{\Delta} - i\omega\ell_N}{2}; \frac32; 
\frac{\rho^2}{\ell_N^2}\right)\;.
\end{align}
For $\omega>0$, these modes are $\delta$-function normalizable and can be superposed into square-integrable wave packets on \eqref{eq:dS2}. Even/odd refers to their parity under $\rho\to -\rho$:
\begin{align}
    f^\text{even}_{\omega \Delta}(t,-\rho)=f^\text{even}_{\omega \Delta}(t,\rho) \;, \qquad f^\text{odd}_{\omega \Delta}(t,-\rho)=-f^\text{odd}_{\omega \Delta}(t,\rho) \;.
\end{align}

\paragraph{Vectors on $dS_2$}

In two dimensions, the field strength $F_{ab}\equiv\partial_a A_b-\partial_b A_a$ has a single independent component, so one may write $F_{ab} =\epsilon_{ab} \,\phi$ for some scalar $\phi$. Here $a,b,c,\dots$ are indices on $dS_2$ and $\epsilon_{t\rho}=\sqrt{-g}\,\tilde\epsilon_{t\rho}=1$ is the Levi-Civita tensor. The Proca equation $\nabla^a F_{ab}=m^2 A_b$ then implies $A_a \propto \epsilon_{ab}\partial^b \phi$. The on-shell condition
\begin{align}\label{eq:dS2Proca}
    \left(-\nabla^2_{dS_2} + \frac{\Delta \bar \Delta+1}{\ell^2_N}\right) A_a =0 \;, \qquad \nabla^a A_a=0 \;, \qquad m^2 \ell^2_N\equiv \Delta (1-\Delta)\equiv \Delta \bar \Delta \;, 
\end{align}
implies that $\phi$ must satisfy \eqref{eq:KGdS2}. A convenient basis of normal modes for the Proca theory on $dS_2$ can therefore be constructed from \eqref{eq:ds2solutions} as
\begin{align}\label{eq:vecfromKG}
    f^\text{even/odd}_{\omega \Delta, a}(t,\rho) \equiv \epsilon_{ab}\partial^b f^\text{even/odd}_{\omega \Delta}(t,\rho) \; .
\end{align}

\paragraph{Transverse traceless tensor on $dS_2$}

Although TT spin-2 harmonics are absent on $S^2$ \cite{rubin1984eigenvalues}, a TT tensor can exist on Lorentzian $dS_2$:\footnote{This mass term is the unique value compatible with the TT conditions.}
\begin{align}\label{eqTTdS2}
    \left(-\nabla^2_{dS_2}   +\frac{2}{\ell_N^2} \right) h_{ab} = 0\;, \quad \nabla^a h_{ab} =0 = h\indices{^a_a} \;. 
\end{align}
This is the wave equation for gravitational perturbations on $dS_2$ in the TT gauge. Every solution is, however, pure gauge: it can be removed by a gauge transformation that preserves the TT conditions. Hence \eqref{eqTTdS2} carries no physical degrees of freedom.

\subsubsection{Pure gauge modes}\label{sec:puregaugenormalmode}

Starting from the normal mode solutions to the Proca equation on the Nariai background \eqref{eq:Nariaimetric} constructed in \cite{Grewal:2022hlo}, one can readily obtain solutions to the ghost system \eqref{eq:ghostsystem}, which parametrizes the residual gauge redundancy of the TT gauge \eqref{eq:EOMTT}.  Below we organize these solutions according to their transformation properties under $SO(d)$. Indices $a,b,c$ and  $i,j,k$ refer to the $dS_2$ and $S^{d-1}$ factors, respectively.

\paragraph{Vector type}

A first class of solutions is
\begin{align}\label{eq:gaugeV}
    \xi_a =0 \;, \qquad \xi_i = f_{\omega} (t,\rho)\, Y^{d-1}_{l,i}\left(\Omega \right) \;, \qquad l\geq 1 \;,
\end{align}
where $Y^{d-1}_{l,i}$ is a transverse vector spherical harmonic on $S^{d-1}$; \eqref{eq:gaugeV} is therefore automatically divergence-free. Substituting into \eqref{eq:ghostsystem} yields an infinite tower of equations of the form \eqref{eq:KGdS2}, with
\begin{align}\label{eq:VmodeDelta}
    \Delta_{1,l} = \frac12 + i \nu_{1,l} \;, \qquad \bar\Delta_{1,l} = \frac12 - i \nu_{1,l} \;, \qquad  \nu_{1,l}=\sqrt{ \frac{l\,(l+d-2)-(d-1)}{d-2}-\frac{1}{4}} \;. 
\end{align}
Hence
\begin{align}
    f_{\omega} (t,\rho) = f^\text{even}_{\omega \Delta_{1,l}}(t,\rho) \quad \text{or} \quad f^\text{odd}_{\omega \Delta_{1,l}}(t,\rho) \;. 
\end{align}

\paragraph{Scalar type I}

The next family is an $SO(1,2)$ vector and an $SO(d)$  scalar:
\begin{align}\label{eq:puregaugeI}
    \xi_a =f_{\omega, a}(t,\rho) Y^{d-1}_{l}\left(\Omega \right) \;, \qquad \xi_i = 0 \;, \qquad l\geq 0 \;. 
\end{align}
This is again automatically transverse. Inserting \eqref{eq:puregaugeI} into \eqref{eq:ghostsystem} produces a tower of Proca equations \eqref{eq:dS2Proca} with
\begin{align}\label{eq:ghostDel0}
    \Delta_{0,l} = \frac12 + i \nu_{0,l} \;, \qquad \bar\Delta_{0,l} = \frac12 - i \nu_{0,l} \;, \qquad\nu_{0,l} =\sqrt{\frac{l\,(l+d-2)-2(d-2)}{d-2} -\frac{1}{4} } \;. 
\end{align}
Thus
\begin{align}
    f_{\omega,a} (t,\rho) = f^\text{even}_{\omega \Delta_{0,l},a}(t,\rho) \quad \text{or} \quad f^\text{odd}_{\omega \Delta_{0,l},a}(t,\rho) \;. 
\end{align}

\paragraph{Scalar type II}

A final scalar-type solution is
\begin{align}\label{eq:puregaugeII}
    \xi_a = \partial_a f_{\omega}(t,\rho) \,Y^{d-1}_{l} \left(\Omega \right)\;, \qquad \xi_i = \frac{\Delta_{0,l} \bar\Delta_{0,l} (d-2)}{l\,(l+d-2)}f_{\omega}(t,\rho) \,\partial_i Y^{d-1}_{l} \left(\Omega \right)\;, \qquad l\geq 1 \;. 
\end{align}
Here $\Delta_{0,l}$ is given in \eqref{eq:ghostDel0}. Proceeding as before, one finds
\begin{align}
    f_{\omega} (t,\rho) = f^\text{even}_{\omega \Delta_{0,l}}(t,\rho) \quad \text{or} \quad f^\text{odd}_{\omega \Delta_{0,l}}(t,\rho) \;.
\end{align}
The relative coefficient in \eqref{eq:puregaugeII} is fixed by the transversality condition.

\subsubsection{Physical transverse traceless modes}

After analyzing the ghost sector, we now present the physical normal mode solutions of the TT-gauge linearized Einstein equation \eqref{eq:EOMTT}. For convenience we write the explicit components of the second-order system:
\begin{align}\label{eq:casimir}
	-\nabla^2  \,h_{ab} +\frac{2}{\ell_N^2}\left( h_{ab} -g_{ab}\, h\indices{^c_c}\right) = 0\;, \quad 
     -\nabla^2 \, h_{ai}  = 0\;, \quad 
     -\nabla^2 \, h_{ij} +\frac{2}{r_N^2} \left(h_{ij} -g_{ij} \, h\indices{^k_k} \right) =0 \;.
\end{align}
As before, we organize the solutions of \eqref{eq:casimir} into tensor, vector, and scalar representations of $SO(d)$. Latin indices $a,b,\dots$ refer to $dS_{2}$; $i,j,\dots$ refer to $S^{d-1}$.

\paragraph{Tensor type}

The simplest physical modes are built from symmetric, transverse–traceless (STT) spin-2 harmonics \eqref{STSH eq} on $S^{d-1}$
\begin{align}\label{eq:normalT}
	h_{ab}=0 \;, \qquad h_{ai}=0 \;, \qquad h_{ij}  = f_{\omega} (t,\rho) \,Y^{d-1}_{l,ij} \left(\Omega \right)\; , \qquad     l\geq 2 \;. 
\end{align}
Because a gauge parameter cannot transform in a rank-2 $SO(d)$ representation,  these modes cannot be gauged away. Substituting \eqref{eq:normalT} into \eqref{eq:casimir} gives
\begin{align}
    f_{\omega} (t,\rho) = f^\text{even}_{\omega \Delta_{2,l}}(t,\rho) \quad \text{or} \quad f^\text{odd}_{\omega \Delta_{2,l}}(t,\rho) \;,
\end{align}
with
\begin{align}
    \Delta_{2,l}=\frac12 + i \nu_{2,l} \;, \qquad \bar\Delta_{2,l}=\frac12 - i \nu_{2,l} \;, \qquad\nu_{2,l}\equiv \sqrt{ \frac{l\,(l+d-2)}{d-2}-\frac14} \;. 
\end{align}

\paragraph{Vector type}

The second type of physical normal modes are constructed from the transverse vector spherical harmonics $Y^{d-1}_i$. One may gauge-fix $h_{ij}$ to zero using \eqref{eq:gaugeV}. On the other hand, $h_{ab}$ does not carry an $S^{d-1}$ vector index. Therefore, the physical vector modes are parametrized by
\begin{align}\label{eq:normalV}
	h_{ab}=0 \;, \qquad h_{ai}=f_{\omega , a}(t,\rho) \,Y^{d-1}_{l,i} \left(\Omega \right)\;, \qquad h_{ij}  =0  \; , \qquad     l\geq 2 \;. 
\end{align}
Inserting \eqref{eq:normalV} into \eqref{eq:casimir} yields
\begin{align}
    f_{\omega,a} (t,\rho) = f^\text{even}_{\omega \Delta_{1,l},a}(t,\rho) \quad \text{or} \quad f^\text{odd}_{\omega \Delta_{1,l},a}(t,\rho) \;,
\end{align}
where $\Delta_{1,l}$ is given in \eqref{eq:VmodeDelta}.

For $l=1$, the would-be vector mode is pure gauge: using the identities
\begin{align}
    f^\text{even}_{\omega \Delta=1, a} &= \epsilon_{ab}\partial^b f^\text{even}_{\omega \Delta=1} = \partial_a \left( -i \omega \ell_N 
 f^\text{odd}_{\omega \Delta=1}\right) \nn\\
    f^\text{odd}_{\omega \Delta=1, a} &= \epsilon_{ab}\partial^b f^\text{odd}_{\omega \Delta=1} = \partial_a \left(-\frac{1}{i \omega \ell_N}  f^\text{even}_{\omega \Delta=1}\right) \;,
\end{align}
one can remove it with a transformation generated by \eqref{eq:gaugeV}. Because $Y^{d-1}_{l=1,i}$ is a Killing vector on $S^{d-1}$, this transformation leaves $h_{ij}$ untouched.

\paragraph{Scalar type} Finally, physical scalar modes are built from the harmonics  $Y^{d-1}_l$. Since $h_{ai}\propto\partial_i Y^{d-1}_{l}$ can always be gauged to zero via a combination of \eqref{eq:puregaugeI} and \eqref{eq:puregaugeII}, we take
\begin{align}\label{eq:normalS}
	h_{ab} &= V^{2,0}_{\omega ,ab} (t,\rho)\, Y_l^{d-1}\left(\Omega \right)\;, \nn\\
    h_{ai }& =0 \;,  \nn\\
	h_{ij}&=\frac{(\Delta_{0,l}+1) (\bar\Delta_{0,l}+1)}{(l-1)(l+d-1)}  f_\omega (t,\rho) V^{d-1,0}_{l,ij}\left(\Omega \right) \;, \quad \qquad l\geq 2\;,
\end{align}
where $V^{d-1,0}_{l,ij}\left(\Omega \right)$ is the transverse (but non-traceless) harmonic \eqref{eq:transHar}, which exists only for $l\neq 1$. We have inserted a relative coefficient for convenience. Recall that $\Delta_{0,l}$ is defined in \eqref{eq:ghostDel0}. The condition $\nabla^\mu h_{\mu a}=0$ implies that $\nabla^a V^{2,0}_{\omega ,ab}=0$. Imposing tracelessness and using the remaining second-order equations \eqref{eq:casimir} fixes
\begin{align}
    V^{2,0}_{\omega ,ab} (t,\rho) = \left(\nabla_{a}\nabla_{b}- \frac{g_{ab}}{2}  \frac{\Delta_{0,l} \bar \Delta_{0,l}}{\ell^2_N}\right)f_\omega (t,\rho)-\frac{1}{2}\frac{(\Delta_{0,l}+1) (\bar\Delta_{0,l}+1)}{\ell^2_N} g_{ab} f_\omega (t,\rho)\;,
\end{align}
and
\begin{align}
    f_{\omega} (t,\rho) = f^\text{even}_{\omega \Delta_{0,l}}(t,\rho) \quad \text{or} \quad f^\text{odd}_{\omega \Delta_{0,l}}(t,\rho) \;.
\end{align}

\subsection{Quasinormal modes and the quasicanonical ideal-gas partition function}\label{sec:QNMZbulk}

For each $l\ge2$, a normal mode solution in any physical sector is captured by the single function
$f^{\text{even/odd}}_{\omega\Delta_{I,l}}(t,\rho)$ defined in \eqref{eq:ds2solutions}.  Recall that
\begin{align}\label{eq:onshellDelta}
     \Delta_{I,l} =1-\bar\Delta_{I,l}= \frac12 + i \nu_{I,l} \;, \qquad  \nu_{I,l}\equiv  \sqrt{\frac{l\,(l+d-2)-(2-I)(d-2+I)}{d-2} -\frac{1}{4} } \;,
\end{align}
where $I=0,1,2$ labeling scalar, vector, and tensor sectors, respectively.

\paragraph{Quasinormal mode spectrum}

Near the horizons $\rho \to \pm \ell_N$, both solutions \eqref{eq:ds2solutions} behave as
\begin{align}
    f^\text{even/odd}_{\omega \Delta} \left(t,\rho\right)\propto e^{-i\omega t}\left[ A^\text{even/odd}_\text{out} \left( \omega\right)  \left(1 - \frac{\rho^2}{\ell_N^2}\right)^{-\frac{i\omega \ell_N}{2}}+A^\text{even/odd}_\text{in} \left( \omega\right)  \left(1 - \frac{\rho^2}{\ell_N^2}\right)^{\frac{i\omega \ell_N}{2}}\right] \;.
\end{align}
The first term corresponds to a wave outgoing towards the horizons, the second to a wave incoming from them, where the outgoing and incoming coefficients are
\begin{alignat}{2}
    A^\text{even}_\text{out}  \left( \omega\right) &=A^\text{even}_\text{in} \left(- \omega\right) &&= \frac{\Gamma\left( i \omega \ell_N\right)}{\Gamma\left( \frac{\Delta+i \omega \ell_N }{2}\right)\Gamma\left( \frac{\bar\Delta+i \omega \ell_N }{2}\right)} \nn\\
   A^\text{odd}_\text{out} \left( \omega\right) &=A^\text{odd}_\text{in} \left(- \omega\right) &&= \frac{\Gamma\left( i \omega \ell_N\right)}{\Gamma\left( \frac{\Delta+1+i \omega \ell_N }{2}\right)\Gamma\left( \frac{\bar\Delta+1+i \omega \ell_N }{2}\right)} \;.
\end{alignat}
For real $\omega$ their ratio is a pure phase, which factorizes as
\begin{align}\label{eq:factorizedS}
    \mathcal{S}^{\text{even/odd}} \left(\omega \right)\equiv \frac{A^\text{even/odd}_\text{out} \left( \omega\right) }{A^\text{even/odd}_\text{in} \left( \omega\right) }=\mathcal{S}^\text{Rin}(\omega)\, \mathcal{S}^{\text{N,even/odd}} \left(\omega \right)  \;.
\end{align}
Upon substituting the on-shell values \eqref{eq:onshellDelta}, the first factor 
\begin{align}\label{eq:rindlerphase}
    \mathcal{S}^\text{Rin}(\omega) = \frac{\Gamma\left(i \omega \ell_N \right)}{\Gamma\left(-i \omega \ell_N \right)} 
\end{align}
is universal to all $l\geq 2$ and only depends on the near-horizon geometry. One way to understand this is that \eqref{eq:rindlerphase} is the scattering phase for the wave equation in the near-horizon Rindler-like region \cite{Law:2022zdq}. On the other hand, the second factor in \eqref{eq:factorizedS}
\begin{align}\label{eq:SN}
    \mathcal{S}^{\text{N,even}}_{I, l}(\omega) = \frac{\Gamma\left( \frac{\Delta_{I,l}-i \omega \ell_N }{2}\right)\Gamma\left( \frac{\bar\Delta_{I,l}-i \omega \ell_N }{2}\right)}{\Gamma\left( \frac{\Delta_{I,l}+i \omega \ell_N }{2}\right)\Gamma\left( \frac{\bar\Delta_{I,l}+i \omega \ell_N }{2}\right)} \;, \quad 
    \mathcal{S}^{\text{N,odd}}_{I, l}(\omega) = \frac{\Gamma\left( \frac{\Delta_{I,l}+1-i \omega \ell_N }{2}\right)\Gamma\left( \frac{\bar\Delta_{I,l}+1-i \omega \ell_N }{2}\right)}{\Gamma\left( \frac{\Delta_{I,l}+1+i \omega \ell_N }{2}\right)\Gamma\left( \frac{\bar\Delta_{I,l}+1+i \omega \ell_N }{2}\right)} \;, 
\end{align}
depends on $l\geq 2$ and the detailed geometry of the Nariai spacetime. Its poles coincide with those of the retarded Green function \cite{Anninos:2011af,Grewal:2024emf}, which we take to define quasinormal modes (QNMs). Explicitly, the physical QNM spectrum is\footnote{The same formula \eqref{eq:QNM} applies to the pure-gauge towers generated by \eqref{eq:gaugeV}, \eqref{eq:puregaugeI}, and \eqref{eq:puregaugeII}.  For those sectors one finds the special values 
\begin{align}
    {\tilde\omega}^{I=1}_{n=0,l=1}=0 \;, \qquad {\tilde\omega}^{I=0}_{n=1,l=0}=0\;, \qquad {\tilde\omega}^{I=0}_{n=0,l=0}=-\frac{1}{\ell_N}
\end{align}
corresponding to QNMs that are undamped or exponentially growing as $t\to \infty$. Such modes could become physical if a timelike boundary were introduced \cite{Anninos:2023epi}.  Clarifying their role may illuminate the Lorentzian origin of the edge partition function discussed in section \ref{sec:PINariai}.}
\begin{empheq}[box=\fbox]{gather}\label{eq:QNM}
    i \omega^I_{nl} = \frac{\Delta_{I,l}+n}{\ell_N} \qquad \text{and} \qquad i {\tilde\omega}^I_{nl} =\frac{\bar\Delta_{I,l}+n}{\ell_N} \;, \qquad l\geq 2\;, \quad n= 0,1, 2, \dots \;, 
\end{empheq}
with degeneracy $D^d_{l,I}$. $I=0,1,2$ denote scalar, vector, and tensor modes, respectively. This spectrum coincides with the QNMs obtained in \cite{Vanzo:2004fy,Lopez-Ortega:2007llk,Lopez-Ortega:2009flo,Venancio:2020ttw} via the Kodama-Ishibashi formalism \cite{Kodama:2003jz}.

\paragraph{Spectral measure}

Through the Krein–Friedel–Lloyd formula, we can define a spectral measure on the continuous normal mode spectrum \cite{Law:2022zdq}
\begin{align}
    \Delta\rho(\omega) = \frac{1}{2\pi i} \sum_{I=0}^2\sum_{l=2}^\infty D^d_{l,I}\partial_\omega \log \left(\frac{\mathcal{S}^{\text{even}}_{I,l} \left(\omega \right)}{\bar{\mathcal{S}}^{\text{even}}_{I,l}(\omega)}\frac{\mathcal{S}^{\text{odd}}_{I,l} \left(\omega \right)}{\bar{\mathcal{S}}^{\text{odd}}_{I,l}(\omega)} \right) \;.
\end{align}
Here $\mathcal{S}^{\text{even/odd}}_{I,l} \left(\omega \right) $ is the scattering phase \eqref{eq:factorizedS} upon substituting \eqref{eq:onshellDelta}. $\bar{\mathcal{S}}^{\text{even/odd}}_{I,l}(\omega)$ is the scattering phase for some reference system. We subtract the universal Rindler phase \cite{Law:2022zdq,Grewal:2022hlo}, that is, we take $\bar{\mathcal{S}}^{\text{even/odd}}_{I,l}(\omega)=\mathcal{S}^\text{Rin}(\omega)$ for any $I=0,1,2$ and $l\geq 2$, so that\footnote{Reference subtraction is standard in contexts such as spectral problems on non-compact hyperbolic space \cite{borthwick_spectral_2016}. A specific reference is often required to ensure, e.g. meromorphic continuation of the resolvent.}
\begin{align}
    \tilde\rho(\omega) = \frac{1}{2\pi i} \sum_{I=0}^2\sum_{l=2}^\infty\partial_\omega \log \left( \mathcal{S}^{\text{N,even}}_{I, l}(\omega)\, \mathcal{S}^{\text{N,odd}}_{I, l}(\omega)\right) \;. 
\end{align}
Employing the digamma series $\psi(x)\equiv \frac{\Gamma'(x)}{\Gamma(x)}=-\gamma +\sum_{n=0}^\infty \left( \frac{1}{n+1}-\frac{1}{n+x}\right)$, and discarding an $(I,l)$-independent infinite constant, one finds
\begin{align}\label{eq:measureQNM}
    \tilde\rho(\omega) = \int_{-\infty}^\infty \frac{dt}{2\pi} \, e^{-i \omega t }\, \chi(t)= \int_{0}^\infty \frac{dt}{2\pi} \left( e^{-i \omega t }+e^{i\omega t}\right)\chi(t)\;,
\end{align}
with
\begin{align}\label{eq:QNMchar}
    \chi(t) \equiv \sum_{I=0}^2\sum_{l=2}^\infty \sum_{n=0}^\infty D^d_{l,I} \, \left( e^{-i \omega^I_{nl}|t| } +e^{-i {\tilde\omega}^I_{nl}|t| }\right)= \sum_{I=0}^2\sum_{l=2}^\infty D^d_{l,I} \, \frac{e^{-\Delta_{I,l}\frac{|t|}{\ell_N}}+e^{-\bar\Delta_{I,l}\frac{|t|}{\ell_N}}}{1-e^{-\frac{|t|}{\ell_N}}} \;. 
\end{align}
In $dS_{d+1}$, the analog of \eqref{eq:QNMchar} is the Harish-Chandra character for the $SO(1,1)$ generator in the dS group $SO(1,d+1)$. In the infinite sum \eqref{eq:QNMchar}, each term with fixed $l\geq 2$ and $I=0,1,2$ is the character of a principal-series representation of $SO(1,2)$. It would be interesting to understand \eqref{eq:QNMchar} as the Harish-Chandra character for the $SO(1,1)$ generator of the full Nariai isometry group $SO(1,2)\times SO(d)$, and to construct Nariai QNMs algebraically, as in \cite{Sun:2020sgn,Ng:2012xp,Jafferis:2013qia,Tanhayi:2014kba}.

\paragraph{Quasicanonical ideal-gas partition function}

Consider an ideal graviton gas at (arbitrary) inverse temperature $\beta>0$. Using the spectral measure \eqref{eq:measureQNM}, we can define a thermal canonical partition function
\begin{align}\label{eq:bulkcanfn}
    \log Z_\text{bulk}\left(\beta \right) = - \int_0^\infty d\omega \, \tilde \rho(\omega)\log \left(e^{\frac{\beta \omega}{2}}-e^{-\frac{\beta \omega}{2}}  \right)=\int_0^\infty \frac{dt}{2t}\frac{1+e^{-\frac{2\pi}{\beta}t}}{1-e^{-\frac{2\pi}{\beta}t}} \chi(t) \;. 
\end{align}
In the second equality we substituted \eqref{eq:measureQNM} and performed the $\omega$-integral, with the $t^{-2}$ pole in the factors multiplying $\chi(t)$ resolved by 
\begin{align}\label{eq:resolve}
    \frac{1}{t^2}\; \to \;\frac{1}{2}\left(\frac{1}{(t+i \epsilon)^2}+\frac{1}{(t-i \epsilon)^2}\right)\;. 
\end{align}
The quantity \eqref{eq:bulkcanfn} is the direct Nariai analog of the dS ``quasicanonical’’ partition function \eqref{introeq:Zbulk}.  We emphasize that all the above results are derived purely in Lorentzian signature on the Nariai spacetime \eqref{eq:Nariaimetric}, with no reference at all to its Euclidean continuation.


\section{1-loop graviton path integrals on closed Einstein manifolds}\label{sec:EinsPI}

We consider the Euclidean path integral for pure Einstein gravity with $\Lambda>0$,
\begin{align}\label{eq:Zgrav}
    \mathcal{Z} = \int \mathcal{D}g\, e^{-S_\text{EH}[g] }\;, \qquad S_\text{EH}[g] = \frac{1}{16\pi G_N} \int d^{d+1} x\sqrt{g} \left( 2\Lambda -R \right)\;.  
\end{align}
For any classical solution $\bar g_{\mu\nu}$ of the Einstein equations,
\begin{align}\label{eq:Riccionshellrelation}
	\bar R = \frac{d+1}{d-1}2\Lambda \;, \qquad {\bar R}_{\mu\nu} = \frac{2\Lambda }{d-1} {\bar g}_{\mu\nu} =\frac{\bar R}{d+1}{\bar g}_{\mu\nu}\;, 
\end{align}
we expand the metric as $g_{\mu\nu}= \bar g_{\mu\nu}+h_{\mu\nu}$ in \eqref{eq:Zgrav} and obtain
\begin{align}
    \mathcal{Z} = \cdots + e^{-S^\text{on-shell}[\bar g] } Z^\text{1-loop}_\text{PI}\left[\mathcal{M}\right]
    \left(1+\cdots \right)+\cdots \;.
\end{align}
For every $d\geq 2$, the leading saddle is the round sphere $S^{d+1}$, whose 1-loop path integral has been studied extensively \cite{Gibbons:1978ji, Christensen:1979iy,Fradkin:1983mq, Allen:1983dg,Taylor:1989ua,GRIFFIN1989295, Mazur:1989ch, Vassilevich:1992rk, Volkov:2000ih,Polchinski:1988ua,Anninos:2020hfj,Law:2020cpj}. Here we study
$Z^\text{1-loop}_\text{PI}\left[\mathcal{M}\right]$ for an arbitrary closed Einstein manifold $\mathcal{M}\neq S^{d+1}$. We will specialize to $\mathcal{M}=S^2\times S^{d-1}$ with $d\geq 3$ in section \ref{sec:PINariai}.

\paragraph{1-loop path integral}

Quadratic graviton fluctuations around a given $\mathcal{M}$ give
\begin{align}\label{eq:1loopgen}
    Z^\text{1-loop}_\text{PI}\left[\mathcal{M}\right] = \int \frac{\mathcal{D}h}{\text{vol} (\mathcal{G})}\, e^{-S^{(2)}_\text{EH}[h] } \;, 
\end{align}
where
\begin{align}\label{eq:linEHsim}
	S^{(2)}_\text{EH}[h] &= \frac{1}{2\mathrm{g}^2}\int_{\mathcal{M}} \bigg( 
	  h^{\mu \nu} \left(-{\bar\nabla}^2\right) h_{\mu \nu}
	+ 2 h^{\mu \nu} {\bar\nabla}_\mu {\bar\nabla}^\rho  h_{\rho \nu} 
	- 2  h\indices{^\lambda_\lambda} {\bar\nabla}^\mu {\bar\nabla}^\nu h_{\mu \nu}
	+    h\indices{^\lambda_\lambda} {\bar\nabla}^2  h\indices{^\rho_\rho} \nn\\
	& \qquad \qquad +\frac{2\Lambda }{ d-1}  \left( h\indices{^\lambda_\lambda}\right)^2   -2 {\bar R}_{\mu\alpha\nu\rho} h^{\mu\nu}h^{\alpha\rho} \bigg) \;. 
\end{align}
Here $\mathrm{g}^2 \equiv 32 \pi G_N$ and we use the shorthand notation: $\int_\mathcal{M} d^{d+1}x\sqrt{g} \to \int_\mathcal{M}$. All covariant derivatives ${\bar\nabla}$ are taken with respect to the background metric $\bar g_{\mu\nu}$; we drop the bars on $\bar{g}_{\mu\nu}$, $\bar\nabla$, and all related geometric quantities from now on.

The functional measure is fixed by
\begin{align}\label{eq:measurenormalization}
	 1=\int \mathcal{D}h\, e^{-\frac{M^2}{2\mathrm{g}^2} (h,h)}  \;. 
\end{align}
Here $M$ is a mass scale inserted to keep the exponent dimensionless. When evaluating the final expressions in some regularization scheme (such as heat kernel \cite{Vassilevich:2003xt}), one could identify $M$ with the appropriate UV regulator. The inner product for spin-$s$ fields $f_{\mu_1 \cdots \mu_s}$ and $g_{\mu_1 \cdots \mu_s}$ is
\begin{align}\label{eq:inner}
    \left( f,g\right)\equiv \int_\mathcal{M} f_{\mu_1 \cdots \mu_s}g^{\mu_1 \cdots \mu_s} \;.
\end{align}
In \eqref{eq:1loopgen}, dividing the measure by the volume $\text{vol} (\mathcal{G})$ of the space of linear diffeomorphisms  removes the overcounting of gauge-equivalent orbits.

\subsection{The geometric approach}

We would like to express \eqref{eq:1loopgen} in terms of determinants of second-order, Laplace-type operators on $\mathcal{M}$. There are multiple ways to proceed—one is the Faddeev–Popov method, as done in \cite{Shi:2025amq}. We will follow the geometric approach \cite{Babelon:1979wd,Mazur:1989by,Bern:1990bh,Vassilevich:1992rk}, which has been applied to massless higher-spin fields on $EAdS_{d+1}$ \cite{Gaberdiel:2010xv,Gaberdiel:2010ar} and $S^{d+1}$ \cite{Law:2020cpj} for any $d\geq 2$. In this framework, we decompose
\begin{align}\label{eq:metricdecom}
	h_{\mu\nu} = h^{TT}_{\mu\nu} + \frac{1}{\sqrt{2}} \left(\nabla_\mu \xi_\nu +\nabla_\nu \xi_\mu  \right) + \frac{g_{\mu\nu}}{\sqrt{d+1}} \tilde  h \;.
\end{align}
The transverse traceless (TT) component satisfies
\begin{align}
	\nabla^\mu h^{TT}_{\mu\nu} = g^{\mu\nu}h^{TT}_{\mu\nu}  =0 \;. 
\end{align}
To make the decomposition \eqref{eq:metricdecom} unique we require the vector $\xi_\mu$ to be orthogonal, with respect to the inner product \eqref{eq:inner}, to every Killing vector of the background:
\begin{align}\label{eq:KVortho}
	\left( \xi,\xi^\text{KV}\right) =0 \;, \qquad \nabla_\mu\xi^\text{KV}_\nu+\nabla_\nu\xi^\text{KV}_\mu=0 \;.
\end{align}
If $\mathcal{M}$ admits conformal Killing vectors (that are not Killing vectors) obeying
$\nabla_\mu\xi^\text{CKV}_\nu+\nabla_\nu\xi^\text{CKV}_\mu=\frac{2}{d+1}g_{\mu\nu}\nabla^\lambda \xi^\text{CKV}_\lambda$, we must also impose $\left( \tilde h,\nabla\cdot \xi^\text{CKV}\right) =0 $. However, this occurs only for $\mathcal{M}=S^{d+1}$,\footnote{Writing $\xi^{\mathrm{CKV}}_\mu = \nabla_\mu \sigma$, the conformal Killing equation on a compact manifold reduces to $\left(-\nabla^{2} - \frac{R}{d}\right)\sigma = 0 $.
By Obata’s theorem \cite{Obata1962CertainCF}, a positively curved Einstein manifold admits a non‑trivial solution of this equation if and only if it is isometric to the round sphere $S^{d+1}$.} so for all other Einstein manifolds no additional condition on $\tilde h$ is necessary.

Under the change of variables \eqref{eq:metricdecom}, the integration measure becomes
\begin{align}\label{eq:measurechange}
	\mathcal{D} h = J \,\mathcal{D} h^{TT} \mathcal{D}'\xi \,\mathcal{D}\tilde h \;. 
\end{align}
Here the prime indicates that Killing vectors are omitted from the $\xi$-integral.  The (flat) measures on the right-hand side are normalized by
\begin{gather}\label{eq:measurenormalizationcomp}
    \int \mathcal{D} h^{TT} \, e^{-\frac{M^2}{2\mathrm{g}^2} \left( h^{TT},h^{TT} \right)  } =\int \mathcal{D} \xi \, e^{-\frac{M^4}{2\mathrm{g}^2} \left( \xi,\xi \right)  }=\int  \mathcal{D}\tilde h \, e^{-\frac{M^2}{2\mathrm{g}^2} \left(\tilde h,\tilde h\right) } =1 \;.
\end{gather}
The Jacobian $J$ will be computed in section \ref{sec:Jacobian}.

\subsubsection{Actions for $h^{TT}$ and $\tilde h$}

Because of the gauge invariance of the linearized Einstein-Hilbert action \eqref{eq:linEHsim}, the pure gauge part in \eqref{eq:metricdecom} drops out, while the TT and scalar parts decouple
\begin{align}\label{eq:actionsplit}
	S\left[h\right] = S\left[h^{TT}\right] +S\left[\tilde h\right]\;.
\end{align}
The TT action is 
\begin{align}\label{eq:genTTaction}
	S\left[h^{TT}\right]  &= \frac{1}{2\mathrm{g}^2}\int_\mathcal{M}  h^{TT,\mu \nu} D^2_{\mu\nu\alpha\beta} h^{TT,\alpha\beta} \;, \qquad D^2_{\mu\nu\alpha\beta}  \equiv - g_{\mu\alpha}g_{\nu\beta}\nabla^2  -2 R_{\mu \alpha \nu \beta } \;.
\end{align}
Path-integrating over $h^{TT}$ gives
\begin{align}\label{eq:ZTT}
    Z^{TT} = \int \mathcal{D}h^{TT} \, e^{-S\left[h^{TT}\right]} \;. 
\end{align}
After some computation, one finds the scalar action
\begin{align}
	S[\tilde h] = -\frac{1}{2\mathrm{g}^2} \frac{d(d-1)}{d+1}\int \tilde h \left( -\nabla^2 -\frac{R}{d}\right) \tilde h \;. 
\end{align}
On any closed manifold the constant mode $\tilde h = f_0=\frac{1}{\sqrt{\text{vol}\left(\mathcal{M}\right)}}$ is present and has a positive action. For non-constant modes, the Lichnerowicz-Obata bound on a closed Einstein manifold $\mathcal{M}$ implies their eigenvalues $\lambda_0$ of $-\nabla^2$ satisfy
\begin{align}
    \lambda_0 \geq \frac{R}{d} \;, \qquad \text{when} \qquad \lambda_0 \neq 0\;, 
\end{align}
with equality if and only if $\mathcal{M}=S^{d+1}$. Thus for $\mathcal{M}\neq S^{d+1}$ all non-constant modes have negative action. To render the path integral convergent we perform the Wick rotation \cite{Gibbons:1978ac}
\begin{align}\label{eq:Wickrotate}
	\tilde{h} \to i^{\mp 1} \, \tilde h \;.
\end{align}
A sign prescription was proposed in \cite{Maldacena:2024spf,Ivo:2025yek}, but we will leave the sign unfixed for now. The constant mode must be rotated back because its action is already positive. This procedure produces a factor
\begin{align}
     i^{\pm 1} \; \prod_{\lambda_0} i^{\mp D_{\lambda_0}} \;. 
\end{align}
The infinite product runs over the full spectrum of $-\nabla^2$ (including the constant mode with $\lambda_0=0$), with $D_{\lambda_{0}}$ the degeneracy of each Laplacian eigenvalue. This ultralocal infinite product can be absorbed into bare couplings, leaving a net phase $i^{\pm1}$. To sum up, 
\begin{align}\label{eq:htildePI}
	Z_{\tilde h} =i^{\pm 1} \int \mathcal{D}\tilde h \, e^{-\frac{1}{2\mathrm{g}^2} \frac{d(d-1)}{d+1}\int \tilde h \left|-\nabla^2 -\frac{R }{d}\right| \tilde h} \;. 
\end{align}

\subsubsection{The group volume factor}\label{sec:GPIdiscussion}

Since the integrations over $\xi_\mu$ are unweighted by the action, the functional integral $\int \mathcal{D}'\xi$ is formally divergent. We choose the volume $\text{vol} (\mathcal{G})$ of the diffeomorphism group in \eqref{eq:1loopgen} so that this divergence is canceled. This is achieved if we take the gauge transformations generated by a vector field $\alpha =\alpha^\mu\partial_\mu$ to act as
\begin{align}\label{eq:gaugetrans}
    \delta_\alpha h_{\mu\nu} = \frac{1}{\sqrt{2}} \left(\nabla_\mu \alpha_\nu +\nabla_\nu \alpha_\mu\right) + O(h) \;. 
\end{align}
This results in a factor
\begin{align}
    \frac{\int \mathcal{D}'\xi}{\text{vol} (\mathcal{G})}= \frac{1}{\text{vol} (G)_{\rm PI}} \;,
\end{align}
where the integral is over only those gauge parameters that leave the background invariant, i.e. the Killing vectors $\xi^\text{KV}_\mu$ of $\mathcal{M}$. With a flat path integral measure and unbounded ranges, $\text{vol} (G)_{\rm PI}$ would be given by a divergent integral 
\begin{align}\label{eq:groupvolumeflat}
    \int \prod_A \frac{M^2}{\sqrt{2\pi}\mathrm{g}}dc_A \; .
\end{align}
Here $c_A$ are the expansion coefficients of a Killing vector $\xi^\text{KV}_\mu = \sum_A c_A f^{KV}_{A,\mu}$ in an orthonormal basis $f^{KV}_{A,\mu}$, which can be viewed as coordinates on the tangent space of the isometry group $G$ near the identity or equivalently the Lie algebra $\mathfrak{g}$ of $G$. We recall that the normalization follows from \eqref{eq:measurenormalizationcomp}.  However, the divergence in \eqref{eq:groupvolumeflat} is simply an artifact of perturbation theory. A natural cure is to replace the perturbative measure \eqref{eq:groupvolumeflat} with the Haar measure $d\mu(c)$ on $G$, and assign
\begin{align}\label{eq:GPI}
    \text{vol} (G)_{\rm PI} = \int_G d\mu(c) \; .
\end{align}
The normalisation of $d\mu$ is chosen its induced bilinear form on $\mathfrak g$ matches the one coming from \eqref{eq:groupvolumeflat}. To evaluate \eqref{eq:GPI}, note that the nonlinear gauge transformations generate an algebra,\footnote{To obtain \eqref{eq:gaugebracket} and \eqref{eq:bracketlie}, one must keep the higher-order terms in \eqref{eq:gaugetrans}: $\delta_\alpha h_{\mu\nu} = \frac{1}{\sqrt{2}} \left(\nabla_\mu \alpha_\nu+\nabla_\nu \alpha_\mu \right) + \frac{1}{\sqrt{2}} \left( \alpha^\rho \nabla_\rho h_{\mu\nu}+\nabla_\mu \alpha^\rho h_{\rho\nu}+\nabla_\nu \alpha^\rho h_{\rho\mu}\right) +O(h^2)$.} with a bracket defined by 
\begin{align}\label{eq:gaugebracket}
    [\delta_\alpha, \delta_{\alpha'}] \equiv \delta_{[[\alpha,\alpha']]} \;. 
\end{align}
Not surprisingly, $[[\cdot ,\cdot]]$ is proportional to the usual Lie bracket between two vector fields
\begin{align}\label{eq:bracketlie}
    [[\alpha, {\alpha'}]] =-\frac{1}{\sqrt{2}}[\alpha, {\alpha'}]_L \;, \qquad [\alpha, {\alpha'}]_L = \left( \alpha^\mu \partial_\mu {\alpha'}^\nu-{\alpha'}^\mu \partial_\mu {\alpha}^\nu\right) \partial_\nu \;.
\end{align}
We stress that the explicit form of the gauge algebra bracket, defined through \eqref{eq:gaugebracket}, depends on the normalization of $h_{\mu\nu}$ and the gauge transformations. 

The Killing vectors, generate a subalgebra of the gauge algebra, with a bracket \eqref{eq:bracketlie} inherited from the latter. This algebra is clearly isomorphic to the Lie algebra $\mathfrak{g}$ of $G$. Suppose there is a known ``canonical" volume $\text{vol} (G)_c $ of $G$,  defined with respect to an inner product in which a standard basis of generators $\xi_c^\text{KV}$ is unit-normalized, $\braket{\xi_c^\text{KV}|\xi_c^\text{KV}}_c = 1$. With the bilinear form induced by the path integral measure, $\xi_c^\text{KV}$ are not normalized to one, but instead to
\begin{align}
    \braket{\xi_c^\text{KV}|\xi_c^\text{KV}}_\text{PI} = \frac{M^4}{2\pi\mathrm{g}^2}\int_\mathcal{M}\xi_c^\text{KV}\cdot\xi_c^\text{KV} \equiv \frac{M^4}{2\pi\mathrm{g}^2}  \norm{\xi_c^\text{KV}}^2 \;.
\end{align}
This implies that the volume \eqref{eq:GPI} is related to $\text{vol} (G)_c $ through
\begin{align}
    \text{vol} (G)_{\rm PI} = \left(\frac{M^2}{\sqrt{2\pi\mathrm{g}^2}}  \norm{\xi_c^\text{KV}} \right)^{\text{dim} G}\text{vol} (G)_c \;. 
\end{align}
In section \ref{sec:1-loopNariai}, we shall evaluate this factor explicitly for $\mathcal{M}=S^{2}\times S^{d-1}$, where  
$G=SO(3)\times SO(d)$.

\subsubsection{The Jacobian}\label{sec:Jacobian}

Consistency between \eqref{eq:measurenormalization} and \eqref{eq:measurenormalizationcomp} requires
\begin{align}\label{eq:Jacobiancondition}
	 1=\int \mathcal{D}h\, e^{-\frac{M^2}{2\mathrm{g}^2} (h,h)} =  
     J\int \mathcal{D} h^{TT} \mathcal{D}'\xi \, \mathcal{D}\tilde h \, e^{-\frac{M^2}{2\mathrm{g}^2} \left( h^{TT}+\sqrt{2}\nabla \xi + \frac{g\tilde h}{\sqrt{d+1}},h^{TT}+\sqrt{2}\nabla \xi + \frac{g\tilde h}{\sqrt{d+1}}\right) } \;.
\end{align}
We have used a compact notation that suppresses the indices on the symmetric tensors and leaves the symmetrization implicit. Specifically, 
\begin{align}
    (\nabla\xi)_{\mu\nu} \equiv \frac{1}{2} \left(\nabla_{\mu} \xi_\nu+ \nabla_{\nu} \xi_\mu\right) \;,
\end{align}
and
\begin{align}
    \left(h^{TT}+\sqrt{2}\nabla\xi+\frac{g}{\sqrt{d+1}}\tilde h\right)_{\mu\nu} \equiv h^{TT}_{\mu\nu} + \frac{1}{\sqrt{2}} \left(\nabla_\mu \xi_\nu +\nabla_\nu \xi_\mu  \right) + \frac{g_{\mu\nu}}{\sqrt{d+1}} \tilde  h \;.
\end{align}
On the right hand side of \eqref{eq:Jacobiancondition}, the TT sector  decouples 
\begin{align}
    \left( h,h\right) = \left( h^{TT},h^{TT}\right)+\left( \sqrt{2}\nabla \xi + \frac{g\tilde h}{\sqrt{d+1}},\sqrt{2}\nabla \xi + \frac{g\tilde h}{\sqrt{d+1}}\right) \;. 
\end{align}
Introduce a shifted scalar field  
\begin{align}
	\tilde h' = \tilde h + \sqrt{\frac{2}{d+1}}\nabla^\lambda \xi_\lambda \;,
\end{align}
which leaves the functional measure unchanged.  Then
\begin{align}
	\left( \sqrt{2}\nabla \xi + \frac{g\tilde h}{\sqrt{d+1}}, \sqrt{2}\nabla \xi + \frac{g\tilde h}{\sqrt{d+1}}\right)  = \left( \tilde h', \tilde h'\right)+\frac12 \left(K\xi,K\xi \right)\;,
\end{align}
where we have defined the differential operator 
\begin{align}
	\left( K\xi\right)_{\mu\nu} \equiv \nabla_\mu\xi_\nu+\nabla_\nu\xi_\mu-\frac{2}{d+1}g_{\mu\nu} \nabla^\lambda  \xi_\lambda \;.
\end{align}
Integrating over $h^{\mathrm{TT}}_{\mu\nu}$ and $\tilde h'$ with the normalizations in \eqref{eq:measurenormalizationcomp} yields
\begin{align}
    \frac{1}{J} =\int  \mathcal{D}'\xi  \, e^{-\frac{M^2}{4\mathrm{g}^2} \left( K\xi ,K\xi \right) } = \det\nolimits' \left(\frac{K^\dagger K}{2M^2}\right)^{-\frac{1}{2}} \;,
\end{align}
where the adjoint $K^\dagger$ is defined with respect to the inner product \eqref{eq:inner}. Decomposing $\xi_\mu=\xi_\mu^{T}+\nabla_\mu\varphi$ and repeating the above steps one finds
\begin{align}
	J = \frac{W_\chi}{Y^T_\xi Y_\sigma} 
\end{align}
with
\begin{align}
	Y^T_\xi  &= \int \mathcal{D}'\xi^T \, e^{-\frac{M^2}{2\mathrm{g}^2} \int {\xi}_T^\mu \left(-\nabla_1^2-\frac{R}{d+1}\right) \xi^T_\mu } \label{eq:ghostPI}\\
	Y_\sigma &= \int \mathcal{D}' \sigma \, e^{-\frac{M^2}{2\mathrm{g}^2}\frac{2d}{d+1} \int \sigma \left( -\nabla^2\right) \left(-\nabla^2 -\frac{R}{d} \right)\sigma} \label{eq:sigmaPI}\\
    	W_\chi & = \int \mathcal{D}' \chi \,  e^{-\frac{M^4}{2\mathrm{g}^2}  \int \chi\left( -\nabla^2\right)\chi} \;. \label{eq:chiPI}
\end{align}
The associated measures are fixed by
\begin{gather}\label{eq:measurenormalizationcomp2}
    \int \mathcal{D}'\xi^T \, e^{-\frac{M^4}{2\mathrm{g}^2} \left( \xi^T,\xi^T \right)  } =\int \mathcal{D}' \sigma \, e^{-\frac{M^6}{2\mathrm{g}^2} \left( \sigma,\sigma \right)  }=\int  \mathcal{D}'\chi \, e^{-\frac{M^6}{2\mathrm{g}^2} \left(\chi,\chi\right) } =1 \;.
\end{gather}
In these expressions, primes indicate that Killing vector modes are omitted from $\xi_\mu^{T}$ and constant modes from $\sigma$ and $\chi$.

\subsection{The result}

Collecting the factors obtained so far we arrive at
\begin{align}
	Z^\text{1-loop}_\text{PI}\left[\mathcal{M}\right] = \frac{1}{\text{vol}\left( G\right)_{\text{PI}} }  \frac{Z^{TT}}{Y^T_\xi}\frac{Z_{\tilde h} W_\chi}{ Y_\sigma} \;.
\end{align}
With the measures \eqref{eq:measurenormalizationcomp} and \eqref{eq:measurenormalizationcomp2}, we can compute each factor. First, for the TT part \eqref{eq:ZTT}, the operator $D^2_{\mu\nu\alpha\beta}  \equiv - g_{\mu\alpha}g_{\nu\beta}\nabla^2  -2 R_{\mu \alpha \nu \beta } $ is bounded from below, but may possess a finite number $N^-_{TT}$ of negative modes. We Wick-rotate them in field space to render their integrals convergent. This results in 
\begin{align}
    Z^{TT} =  
    i^{\mp N^-_{TT}} \det \left| \frac{D^2}{M^2} \right|^{-\frac12} \;.
\end{align}
Second, the ghost kinetic operator $-\nabla_1^2-\frac{R}{d+1}$ is non-negative;\footnote{See e.g. section 3.2 of \cite{Shi:2025amq} or appendix A.2 of \cite{Ivo:2025yek}.} its zero modes are precisely the Killing vectors and are excluded from \eqref{eq:ghostPI}. Thus,
\begin{align}\label{eq:Zxi}
    Y^T_\xi = \det\nolimits' \left( \frac{-\nabla_1^2-\frac{R}{d+1}}{M^2}\right)^{-\frac{1}{2}}\;.
\end{align}
Third, the scalar contributions \eqref{eq:ghostPI}, \eqref{eq:sigmaPI} and \eqref{eq:chiPI} largely cancel
\begin{align}
    \frac{Z_{\tilde h} W_\chi}{ Y_\sigma} &=i^{\pm 1}   \int \frac{dc}{\sqrt{2\pi }\mathrm{g}} \, e^{-\frac{1}{2\mathrm{g}^2} \frac{d-1}{d+1}R \, c^2}  \int \mathcal{D}'\tilde h \, e^{-\frac{1}{2\mathrm{g}^2}\frac{d-1}{2}\int \tilde h ^2} \nn\\
    &= i^{\pm 1} \left( \frac{d+1}{d-1}\frac{M^2}{R }\right)^{\frac12}{\prod_{\lambda_0}}' \left(\frac{2}{d-1}\right)^\frac{D_{\lambda_0}}{2}\nn\\
    &= i^{\pm 1} \left( \frac{d-1}{4\Lambda}M^2\right)^{\frac12} 
\end{align}
On the first line, the first factor corresponds to the integration over the constant mode $f_0=\frac{1}{\sqrt{\text{vol}\left(\mathcal{M}\right)}}$ (with coefficient $c$) in \eqref{eq:htildePI}, with factor $\frac{1}{\sqrt{2\pi}\mathrm{g}}$ fixed by the normalization condition \eqref{eq:measurenormalizationcomp}. The non-constant mode integrations in \eqref{eq:htildePI} combine with $Y_\sigma$ and $W_\chi$ to give the second factor. On the last line, we absorbed an ultralocal infinite constant to local coupling constants, and used \eqref{eq:Riccionshellrelation} to express the Ricci scalar in terms of $\Lambda$.

Putting these pieces together, the 1-loop graviton path integral on any ($d+1$)-dimensional connected, closed Einstein manifold $\mathcal{M}\neq S^{d+1}$ with $\Lambda>0$ is  
\begin{empheq}[box=\fbox]{align}\label{eq:ZPIgenEinstein}
    Z^\text{1-loop}_\text{PI} \left[\mathcal{M} \right]= \frac{i^{\mp N^-_{TT}\pm 1}}{\text{vol}\left( G\right)_c} \left(\frac{M^2}{\sqrt{2\pi\mathrm{g}^2}}  \norm{\xi_c^\text{KV}} \right)^{-\text{dim } G} \left( \frac{d-1}{4\Lambda}M^2\right)^{\frac12}  \frac{\det\nolimits' \left( \frac{-\nabla_1^2-\frac{R}{d+1}}{M^2}\right)^{\frac{1}{2}}}{\det \left| \frac{D^2}{M^2} \right|^{\frac12}}\;.
\end{empheq}
The sign in $i^{\pm N^{-}_{\mathrm{TT}}\pm1}$ has not yet been fixed. Following the prescription of \cite{Maldacena:2024spf,Ivo:2025yek}—namely, taking the Newton constant slightly complex, $\frac{1}{G_{N}}=\frac{1}{|G_{N}|}(1-i\epsilon)$—selects the minus sign in both the TT-sector Wick rotation \eqref{eq:ZTT} and the scalar rotation \eqref{eq:Wickrotate}, giving the definitive phase  
\begin{align}\label{eq:fixedphase}
  i^{-N^{-}_{\mathrm{TT}}+1}.
\end{align}


\section{Linearized gravity on $S^2\times S^{d-1}$}\label{sec:PINariai}

In this section we study the Euclidean path integral for pure Einstein gravity with $\Lambda>0$ expanded around the Euclidean Nariai geometry $S^2\times S^{d-1}$ for any $d\geq 3$.

\subsection{1-loop graviton Nariai path integral}\label{sec:1-loopNariai}

\paragraph{The Euclidean Nariai geometry}

Starting from the Lorentzian Nariai spacetime \eqref{eq:Nariaimetric} and performing the analytic continuation
\begin{align}
    t \to -i\tau \;, \qquad \tau \simeq \tau + 2\pi \ell_N \;, 
\end{align}
we arrive at an Einstein metric on $S^2\times S^{d-1}$:
\begin{align}\label{eq:ENariaimetric}
ds^2=\left(1-\frac{\rho^2}{\ell_N^2}\right)d \tau^2 +\frac{d\rho^2}{1-\frac{\rho^2}{\ell_N^2}}+r_N^2 \, d\Omega_{d-1}^2\;, \qquad -\ell_N < \rho <\ell_N \; .
\end{align}
In other words, the geometry is the direct product of a round $S^{2}$ of radius $\ell_{N}$, and a round $S^{d-1}$ of radius $r_{N}$. Its isometry group is $SO(3)\times SO(d)$.  The only non-vanishing components of the Riemann tensor are those with all indices in the $S^{2}$ factor or all in the $S^{d-1}$ factor:
\begin{align}\label{eq:RiemNariai}
		R_{abcd} = \frac{g_{ac}g_{bd}-g_{bc}g_{ad}}{\ell_N^2} \;, \qquad 
		R_{ijkl}  = \frac{g_{ik}g_{jl}-g_{jk}g_{il}}{r_N^2} \;.
\end{align}
Here $a,b,c,d$ are $S^{2}$ indices and $i,j,k,l$ are $S^{d-1}$ indices. The Ricci scalar is
\begin{align}\label{eq:RicciscalarNariai}
	R = \frac{2}{\ell^2_N} + \frac{(d-1)(d-2)}{r^2_N} =  \frac{(d+1)(d-2)}{r^2_N} = \frac{d+1}{\ell_N^2}\;. 
\end{align}
The on-shell action reads
\begin{align}\label{eq:Nariaionshell}
    S^\text{on-shell}_{S^2\times S^{d-1}}[g^*] = - \frac{A}{4G_N} \;, \qquad A =2 \times r^{d-1}_N\text{vol}\left( S^{d-1}\right) \;,
\end{align}
where $A$ is the total area of the black hole and cosmological horizon of the Nariai spacetime \eqref{eq:Nariaimetric}.

In any $d\geq 4$, $S^2\times S^{d-1}$ is the first subleading saddle after $S^{d+1}$. For $d=3$, $S^2\times S^2$ is less dominant than $\mathbb{CP}^2$, which has a larger isometry group $SU(3)$ \cite{Anninos:2025ltd}.

\subsubsection{The 1-loop path integral}

We now analyze the 1-loop path integral \eqref{eq:1loopgen} around the saddle \eqref{eq:ENariaimetric}, in parallel with the round $S^{d+1}$ calculation of \cite{Anninos:2020hfj,Law:2025ktz}. Using the general formula \eqref{eq:ZPIgenEinstein} with $\mathcal{M}=S^{2}\times S^{d-1}$ we obtain
\begin{empheq}[box=\fbox]{align}\label{eq:NariaiZPI}
	 Z^\text{1-loop}_\text{PI} \left[S^2 \times S^{d-1} \right] 
    &=  \frac{1}{\text{vol}\left( SO(3)\times SO(d)\right)_{\text{PI}}}\left( \frac{d-1}{4\Lambda}M^2\right)^{\frac12} \frac{\det\nolimits'\left( \frac{-\nabla_1^2-\frac{R}{d+1}}{M^2}\right)^\frac12}{\det\left| \frac{D^2}{M^2}\right|^{\frac12}}  \;.
\end{empheq}
We recall that in the ratio of determinants, the operator
\begin{align}\label{eq:TTLaplacian}
    D^2_{\mu\nu\alpha\beta}  \equiv - g_{\mu\alpha}g_{\nu\beta}\nabla^2  -2 R_{\mu \alpha \nu \beta } 
\end{align}
is the kinetic operator acting on the transverse-traceless (TT) part of $h_{\mu\nu}$. As shown in appendix \ref{sec:spec}, \eqref{eq:TTLaplacian} possesses a single negative mode, so $N^-_{TT}=1$. We have taken the phase prescription \eqref{eq:fixedphase}. We also recall that the prime on $\det'$ indicates that Killing-vector zero modes are removed from  the ghost determinant.

\paragraph{Isometry group volume}

For $S^2\times S^{d-1}$ the isometry group is $SO(3)\times SO(d)$. The normalized generators
\begin{align}\label{eq:normalizedgenerators}
    J_{AB} = - \sqrt{2} \left(X_A\partial_{X^B} - X_B \partial_{X^A} \right) \;, \qquad M_{IJ} = - \sqrt{2} \left(Y_I\partial_{Y^J} - Y_J \partial_{Y^I} \right)
\end{align}
satisfy the standard $\mathfrak{so}(3)\oplus \mathfrak{so}(d)$ commutation relations under the bracket \eqref{eq:bracketlie}:
\begin{gather}
	\left[ \left[ J_{AB},J_{CD}\right] \right] = \delta_{BC} J_{AD} -\delta_{BD} J_{AC} -\delta_{AC} J_{BD} +\delta_{AD} J_{BC} \nn\\
	\left[ \left[ M_{IJ},M_{KL}\right] \right] = \delta_{JK} M_{IL} -\delta_{JL} M_{IK} -\delta_{IK} M_{JL} +\delta_{IL} M_{JK} \nn\\
	\left[ \left[ J_{AB},M_{IJ}\right] \right] = 0
\end{gather}
In these expressions, $A,B,C,D=1,2,3$ and $I,J,K,L=1,\dots ,d$. $X^A$ and $Y^I$ are the embedding coordinates of the Euclidean Nariai geometry in $\mathbb{R}^{d+3}=\mathbb{R}^3 \times \mathbb{R}^d$:
\begin{align}
    X_1^2 +X_2^2 +X_3^2=\ell_N^2 \;, \qquad  Y_1^2 +Y_2^2 +\cdots + Y_{d}^2 = r_N^2\;.
\end{align}
The ``canonical" volume of $SO(3)\times SO(d)$, 
\begin{align}
    \text{vol}\left( SO(3)\times SO(d)\right)_{\text{c}} 
    =\left( \prod_{A=1}^2 \text{vol} \left( S^A\right) \right)\left( \prod_{I=1}^{d-1} \text{vol} \left( S^I\right) \right) = 8\pi^2  \prod_{I=1}^{d-1} \text{vol} \left( S^I\right) \;, 
\end{align}
is measured with respect to a metric where the generators \eqref{eq:normalizedgenerators} are normalized to unity. Here $\text{vol} \left( S^n\right)=\frac{2\pi^{\frac{n+1}{2}}}{\Gamma\left(\frac{n+1}{2}\right)} $ denotes the volume of a unit round $S^n$. As explained in section \ref{sec:GPIdiscussion}, such a ``canonical" metric does not coincide with that induced by the path integral measure -- the latter depends on the coupling constants in the theory and normalization conventions. In our case, with the bilinear form induced by the path integral measure \eqref{eq:groupvolumeflat}, we can compute
\begin{align}
	\braket{J_{12}|J_{12}}_\text{PI} &=\frac{M^4}{2\pi \mathrm{g}^2} \int_{\mathcal{M}} \left( J_{12}\right)^{A} \left( J_{12}\right)_{A} 
    = \frac{2\ell_N^2}{16\pi G}\frac{2\ell_N^2 r_N^{d-1}}{3} \text{vol} \left( S^{d-1}\right)  M^4\nn\\
	\braket{M_{12}|M_{12}}_\text{PI} &=\frac{M^4}{2\pi \mathrm{g}^2} \int_{\mathcal{M}} \left( M_{12}\right)^{I} \left( M_{12}\right)_{I} = \frac{2r_N^2}{16\pi G}\frac{2\ell_N^2 r_N^{d-1}}{d} \text{vol} \left( S^{d-1}\right) M^4 \;.
\end{align}
We then conclude that 
\begin{align}\label{eq:NariaiGPI}
	\text{vol}\left( SO(3)\times SO(d)\right)_{\text{PI}} 
	=& \left( \frac{A}{8\pi G}\frac{\ell_N^4 M^4}{3} \right)^\frac32  \left( \frac{A}{8\pi G}\frac{r_N^2\ell_N^2 M^4}{d}   \right)^\frac{d(d-1)}{4}\text{vol}\left( SO(3)\times SO(d)\right)_{\text{c}} \;. 
\end{align}
Here $A$ is the total area \eqref{eq:Nariaionshell} of the cosmological and black hole horizons. Taking \eqref{eq:NariaiGPI} into account, \eqref{eq:NariaiZPI} recovers the $d=3$ result in \cite{Volkov:2000ih} (up to the phase resulting from \eqref{eq:Wickrotate}, which was discarded there).

\subsection{From heat kernel to the bulk-edge split}

\subsubsection{Spectra of Laplacians on $S^2 \times S^{d-1}$}\label{sec:spectra}

To proceed we must obtain the full spectra of the Laplacians appearing in \eqref{eq:NariaiZPI}. This was worked out for $d=3$ in \cite{Volkov:2000ih}; we generalize the analysis in appendix \ref{sec:spec}, where the explicit eigenfunctions are derived. Below we merely summarize the eigenvalues and degeneracies. Analogously to the Lorentzian normal modes of section \ref{sec:normalmode}, we organize the spectra according to $SO(d)$ UIRs.

\paragraph{The ghost Laplacian}

We summarize the eigenvalues and degeneracies of $-\nabla_1^2-\frac{R}{d+1}$ as follows:
\begin{itemize}
    \item Vector type:
    \begin{align}\label{eq:NeigvalVghost}
	{\tilde \lambda}^{(1)}_{L,l} = \frac{L(L+1)-1}{\ell_N^2}+\frac{l\,(l+d-2)-1}{r_N^2} \;, \qquad L \geq 1\;, \quad l\geq 1 \;,
\end{align}
degeneracy $D^3_{L,0} D_{l,1}^{d}$.

    \item Scalar type:
    \begin{align}\label{eq:NeigvalS}
	{\tilde \lambda}^{(0)}_{L,l} = \frac{L(L+1)-2}{\ell_N^2}+\frac{l\,(l+d-2)}{r_N^2}
    \qquad 
    \begin{cases}
        L \geq 1 \;, \quad l\geq 0 \\
        L \geq 1 \;, \quad l\geq 1
    \end{cases} \;,
\end{align}
degeneracy $D^3_{L,0} D^d_{l,0}$.
\end{itemize}
These may be written uniformly  as 
\begin{empheq}[box=\fbox]{align}\label{eq:uniformeigenghost}
    {\tilde\lambda}^{(I)}_{L,l} =\frac{\left(L+\frac{1}{2}\right)^2+\nu_{I,l}^2}{\ell_N^2} \;, \quad \nu_{I,l}\equiv  \sqrt{\frac{l\,(l+d-2)-(2-I)(d-2+I)}{d-2} -\frac{1}{4} } \;, \quad I=0,1 \;.
\end{empheq}

\paragraph{The transverse-traceless Laplacian}

We summarize the eigenvalues and degeneracies of the TT Laplacian \eqref{eq:TTLaplacian} as follows: 
\begin{itemize}
    \item Tensor type:
\begin{align}\label{eq:NeigvalT}
	\lambda^{(2)}_{L,l} = \frac{L(L+1)}{\ell_N^2}+\frac{l\,(l+d-2)}{r_N^2} \;, \qquad L \geq 0 \;, \quad l\geq 2 \;,
\end{align}
degeneracy $D^3_{L,0} D_{l,2}^{d}$.

    \item Vector type:
\begin{align}\label{eq:NeigvalV}
	\lambda^{(1)}_{L,l} = \frac{L(L+1)-1}{\ell_N^2}+\frac{l\,(l+d-2)-1}{r_N^2} \;,
    \qquad 
    \begin{cases}
        L \geq 1 \;, \quad l\geq 0 \\
        L \geq 1 \;, \quad l\geq 1
    \end{cases} \;,
\end{align}
degeneracy $D^3_{L,0} D_{l,1}^{d}$. 

    \item Scalar type:
    \begin{align}\label{eq:NeigvalTS}
	\lambda^{(0)}_{L,l} = \frac{L(L+1)-2}{\ell_N^2}+\frac{l\,(l+d-2)}{r_N^2} \;,
    \begin{cases}
        L \geq 2 \;, \quad l\geq 1 \\
        L \geq 2 \;, \quad l\geq 2 \\
        L \geq 0 \;, \quad l\geq 0 \;, \text{ except } (L,l) = (0,1),(1,0) \text{ or } (1,1) 
    \end{cases} ,
\end{align}
degeneracy $D^3_{L,0} D^d_{l,0}$. The mode $(L,l)=(0,0)$ has the negative eigenvalue $\lambda^{(0)}_{0,0} = -\frac{4\Lambda}{d-1}$.

\end{itemize}
We can express \eqref{eq:NeigvalT}, \eqref{eq:NeigvalV} and \eqref{eq:NeigvalS} uniformly as
\begin{empheq}[box=\fbox]{align}\label{eq:uniformeigen}
    \lambda^{(I)}_{L,l} =\frac{\left(L+\frac{1}{2}\right)^2+\nu_{I,l}^2}{\ell_N^2} \;, \quad \nu_{I,l}\equiv  \sqrt{\frac{l\,(l+d-2)-(2-I)(d-2+I)}{d-2} -\frac{1}{4} } \;, \quad I=0,1,2 \;. 
\end{empheq}

Comparing \eqref{eq:uniformeigenghost} with \eqref{eq:uniformeigen}, we find the isospectral relations
\begin{empheq}[box=\fbox]{align}\label{eq:isospectral}
    \lambda^{(1)}_{L,l}  ={\tilde \lambda}^{(1)}_{L,l} \;, \qquad \lambda^{(0)}_{L,l} = {\tilde \lambda}^{(0)}_{L,l} \;.
\end{empheq}

\subsubsection{Bulk-edge split of $Z_\text{PI}$}

In heat kernel regularization, \eqref{eq:NariaiZPI} becomes
\begin{align}\label{eq:ZPIheat}
	 \log Z_\text{PI}  
	&= \log  \frac{ 1}{\text{vol}\left( SO(3)\times SO(d)\right)_{\text{PI}}} 
    + \int_0^\infty \frac{d\tau}{2\tau} e^{-\frac{\epsilon^2}{4\tau}} K(\tau)\;,
\end{align}
with
\begin{align}\label{eq:heatkerneldef}
	K(\tau)= \Tr \, e^{-\left| D^2\right| \tau} + e^{-\left| \lambda^{(0)}_{0,0}\right| \tau}- \Tr' \, e^{-\left( -\nabla_1^2-\frac{R}{d+1}\right) \tau} \;.
\end{align}
Comparing the contribution from a single mode on both sides: $\int_0^\infty \frac{d\tau}{2\tau} e^{-\frac{\epsilon^2}{4\tau}} e^{-\lambda \tau}=K_0\left(\epsilon\sqrt{ \lambda}\right)\to -\frac{1}{2}\log \frac{\lambda}{M^2}$, we identify the UV regulator $\epsilon\to 0$ with the mass scale through $M = \frac{2e^{-\gamma}}{\epsilon}$, where $\gamma$ is the Euler–Mascheroni constant. The term $e^{-\left| \lambda^{(0)}_{0,0}\right| \tau}$ originates from the factor  $\left( \frac{d-1}{4\Lambda}M^2\right)^{\frac12}$ in \eqref{eq:NariaiZPI}.

With the spectra \eqref{eq:uniformeigenghost} and \eqref{eq:uniformeigen}, we immediately write 
\begin{align}\label{eq:phyhk}
	\Tr \, e^{-\left| D^2\right| \tau}  & =  \sum_{L=0}^\infty\sum_{l=2}^\infty D^3_{L,0} D_{l,2}^{d} \, e^{-\lambda^{(2)}_{L,l} \tau } + \sum_{L=1}^\infty \left( \sum_{l=2}^\infty+\sum_{l=1}^\infty\right)  D^3_{L,0} D_{l,1}^{d} \, e^{-\lambda^{(1)}_{L,l} \tau }  \nn\\
	& \qquad + \left( \sum_{L=0}^\infty\sum_{l=2}^\infty+\sum_{L=2}^\infty\sum_{l=1}^\infty+\sum_{L=2}^\infty\sum_{l=0}^\infty\right) D^3_{L,0} D^d_{l,0} \, e^{-\lambda^{(0)}_{L,l} \tau } + e^{-\left| \lambda^{(0)}_{0,0}\right| \tau} 
\end{align}
and 
\begin{align}\label{eq:ghosthk}
	\Tr' \, e^{-\left( -\nabla_1^2-\frac{R}{d+1}\right) \tau} & = \sum_{L=0}^\infty \sum_{l=1}^\infty D^3_{L,0} D_{l,1}^{d} \, e^{-{\tilde\lambda}^{(1)}_{L,l} \tau }  - D^{d}_{1,1} \, e^{-{\tilde\lambda}^{(1)}_{0,1} \tau }\nn\\
	& \qquad + \sum_{L=1}^\infty \left( \sum_{l=0}^\infty+\sum_{l=1}^\infty\right)  D^3_{L,0} D^d_{l,0} \, e^{-{\tilde \lambda}^{(0)}_{L,l} \tau } -D^3_{1,0} \, e^{-{\tilde \lambda}^{(0)}_{1,0} \tau } \;. 
\end{align}
Thanks to the isospectrality \eqref{eq:isospectral}, we find substantial cancellations in \eqref{eq:heatkerneldef}, yielding
\begin{align}\label{eq:heatsumresult}
	K(\tau) 
    &=K_\text{bulk}(\tau) + K_\text{edge}(\tau) \;, \nn\\
   K_\text{bulk}(\tau) &=  \sum_{L=0}^\infty D^3_{L,0} \sum_{I=0}^2\sum_{l=2}^\infty  D_{l,I}^{d} \, e^{-\lambda^{(I)}_{L,l} \tau } \;,  \nn\\
   K_\text{edge}(\tau)&=-2 \left(\sideset{}{'}\sum_{l=-1}^\infty D_{l,1}^{d} \, e^{-|\lambda^{(1)}_{0,l} |\tau }   +D^3_{1,0}\sum_{l=1}^\infty  D^d_{l,0} \, e^{-\lambda^{(0)}_{1,l} \tau } \right)\;, 
\end{align}
where $\sideset{}{'}\sum_{l=-1}^\infty$ omits $l=1$ (for which $\lambda^{(1)}_{0,l=1}=0$) and we have used
\begin{align}
	D^d_{0,1}=0\;, \qquad D_{-1,1}^{d} =-1 \;, \qquad  \lambda^{(0)}_{0,0} =  \lambda^{(1)}_{0,-1}
\end{align}
to extend the sum to $l=-1$.

\paragraph{Bulk partition function}

The bulk contribution to the heat kernel gives
\begin{align}\label{eq:ZbulkPI}
    \log Z_\text{bulk} = \int_0^\infty \frac{d\tau}{2\tau} e^{-\frac{\epsilon^2}{4\tau}} K_\text{bulk}(\tau) \;. 
\end{align}
This expression is local, in the sense that it does not contain logarithmic divergence for even $d$. Recall that a logarithmic divergence arises from a $O(\tau^{0})$ term in the small-$\tau$ asymptotic expansion of the heat kernel
\begin{align}\label{eq:Kbulk}
    K_\text{bulk}(\tau) = \left(\sum_{L=0}^\infty D^3_{L,0} \, e^{-\frac{\left(L+\frac{1}{2}\right)^2}{\ell_N^2} \tau }\right)\left(\frac{1}{2}\sum_{I=0}^2\sum_{l=-\infty}^\infty  D_{l,I}^{d} \, e^{-\frac{\nu_{I,l}^2}{\ell_N^2} \tau }\right) \;,
\end{align}
where we have used $D_{l,I}^{d}=-D_{I-1,l+1}^{d}$, and $D_{l,I}^{d}=D_{-(d-2)-l,I}^{d}$ when $d$ is even, to extend the $l$-sum to all integers. Applying the Euler-Maclaurin formula, one can show that the small-$\tau$ asymptotic expansion of the $(l,I)$-sum has only half-integer powers of $\tau$,
and the $L$-sum generates only integer powers \cite{Kluth:2019vkg}, implying the absence of the $O(\tau^0)$ term.

We now insert \eqref{eq:uniformeigen} and employ the Hubbard-Stratonovich trick for the $L$-sum \cite{Anninos:2020hfj},
\begin{equation}
	\sum_{L=0}^\infty D_{L}^{3}\, e^{-\frac{\tau}{\ell^{2}_N}\left(L+\frac12\right)^2} = \int_{\mathbb{R} + i\delta} du \ \frac{e^{-\frac{u^2}{4\tau}}}{\sqrt{4\pi \tau}} \frac{1+e^{i\frac{u}{\ell_N}}}{1-e^{i\frac{u}{\ell_N}}}  \frac{e^{i\frac{u}{2\ell_N}}}{1-e^{i\frac{u}{\ell_N}}} \; ,
\end{equation}
with $\epsilon>\delta >0$. Performing the $\tau$-integral yields
\begin{align}
	\log Z_{\rm bulk} = \int_{ \mathbb{R} + i\delta}  \frac{du}{2\sqrt{u^2+\epsilon^2}}  \frac{1+e^{i\frac{u}{\ell_N}}}{1-e^{i\frac{u}{\ell_N}}} \sum_{I=0}^2\sum_{l=2}^\infty D^{d}_{l,I}  \frac{e^{i\frac{u}{2\ell_N}-\frac{\nu_{I,l}}{\ell_N} \sqrt{u^2+\epsilon^2}}}{1-e^{i\frac{u}{\ell_N}}}  \;. 
\end{align}
We fold the contour along the two sides of the branch cut from $u = i \epsilon$ to $u=i \infty$. Changing variables $u=it$ and using that the square root takes opposite signs on both sides of the cut, we transform this to an integral
\begin{align}\label{eq:PIbulk}
     \log Z_{\rm bulk} &= \int_\epsilon^\infty \frac{dt}{2\sqrt{t^2-\epsilon^2}}\frac{1+e^{-\frac{t}{\ell_N}}}{1-e^{-\frac{t}{\ell_N}}}\chi_\epsilon (t)
\end{align}
with
\begin{align}\label{eq:regulatedchi}
    \chi_\epsilon (t) = \sum_{I=0}^2\sum_{l=2}^\infty  D_{l,I}^{d} \frac{e^{-\frac{t}{2\ell_N}-i \frac{\nu_{I,l}}{\ell_N}\sqrt{t^2-\epsilon^2}}+e^{-\frac{t}{2\ell_N}+i \frac{\nu_{I,l}}{\ell_N}\sqrt{t^2-\epsilon^2}}}{1-e^{-\frac{t}{\ell_N}}} \;.
\end{align}
Setting $\epsilon\to0$ recovers the quasicanonical ideal-gas partition function \eqref{eq:bulkcanfn} at the inverse Nariai temperature $\beta=\beta_{N}$.

\paragraph{Edge partition function}

Let us now turn to 
\begin{align}\label{eq:Zedge}
    &\quad \; \log Z_{\rm edge} \nn\\
    &\equiv \log  \frac{1}{\text{vol}\left( SO(3)\times SO(d)\right)_{\text{PI}}} 
    + \int_0^\infty \frac{d\tau}{2\tau} e^{-\frac{\epsilon^2}{4\tau}} K_{\rm edge} (\tau) \nn\\
    &= \log  \frac{1}{\text{vol}\left( SO(3)\times SO(d)\right)_{\text{PI}}} - 2 \int_0^\infty \frac{d\tau}{2\tau} e^{-\frac{\epsilon^2}{4\tau}}\left(\sideset{}{'}\sum_{l=-1}^\infty D_{l,1}^{d} \, e^{-\left|\lambda^{(1)}_{0,l} \right|\tau }   +D^3_{1,0}\sum_{l=1}^\infty  D^d_{l,0} \, e^{-\lambda^{(0)}_{1,l} \tau } \right)\;. 
\end{align}
With the explicit expressions for the eigenvalues, 
\begin{align}
	\lambda^{(1)}_{0,l}=\frac{l\,(l+d-2)-1-(d-2)}{r_N^2}\;, \qquad 	\lambda^{(0)}_{1,l}=\frac{l\,(l+d-2)}{r_N^2} \;, 
\end{align}
this can be rewritten in terms of determinants on $S^{d-1}$:
\begin{align}\label{eq:Zedgedet}
    Z_{\rm edge} &= \frac{1}{\text{vol}\left( SO(3)\times SO(d)\right)_{\text{PI}}} \left[ \det_{-1}\nolimits' \left| \frac{-\nabla_1^2-\frac{d-2}{r^2_N}}{M^2}\right|^\frac{1}{2}\det\nolimits' \left(\frac{-\nabla_0^2}{M^2}\right)^{\frac32} \right]^2
\end{align}
where $-\nabla_1^2$ and $-\nabla_0^2$ are, respectively, the Laplacians on transverse vectors and on scalars on $S^{d-1}$. Recall that the group volume is given by \eqref{eq:NariaiGPI}.

\paragraph{Log-coefficients for even $d+1$}

Up to now we have established
\begin{align}\label{eq:ZPIsofar}
    Z^\text{1-loop}_\text{PI} \left[S^2 \times S^{d-1} \right] = Z_\text{bulk} \left( \beta = \beta_N\right) Z_\text{edge} \;,
\end{align}
with $Z_{\text{bulk}}$ given in \eqref{eq:PIbulk} and $Z_{\text{edge}}$ in \eqref{eq:Zedgedet}.   Since $Z_\text{bulk}$ is a local expression, the factorization \eqref{eq:ZPIsofar} means that $Z_\text{edge}$ is local as well. One can verify this directly by applying Euler-Maclaurin formula to \eqref{eq:Zedge}. Alternatively, following \cite{Anninos:2020hfj}, \eqref{eq:Zedgedet} can be written in terms of integrals involving $SO(1,d-1)$ characters; the log-extraction recipe of appendix C in \cite{Anninos:2020hfj} again yields a vanishing
$\log \frac{\ell_N}{\epsilon}$ coefficient. In both checks one must include the group volume contribution \eqref{eq:NariaiGPI} with the identification $M = \frac{2e^{-\gamma}}{\epsilon}$.

For odd $d$ a logarithmically divergent term does appear:
\begin{align}
    \log Z = \cdots + \alpha \log \frac{\ell_N}{\epsilon} + O\left(\epsilon^0\right)
\end{align}
where $Z$ denotes any factor in \eqref{eq:ZPIsofar}, and $\cdots$ denotes the more divergent terms. We tabulate the log-coefficients for various odd $d$ in Table \ref{tab:log}.
\begin{table}[H]
\centering
\renewcommand{\arraystretch}{1.5} 
\begin{tabular}{|c|c|c|c|}
\hline
$d$ & $\alpha_{\mathrm{bulk}}$ & $\alpha_{\mathrm{edge}}$ & $\alpha_{\mathrm{PI}}=\alpha_{\mathrm{bulk}}+\alpha_{\mathrm{edge}}$ \\
\hline
3  & $\frac{22}{45}$                 & $-\frac{8}{3}$           & $-\frac{98}{45}$                  \\
\hline
5  & $-\frac{509}{630}$                 & $-\frac{36}{5}$           & $-\frac{1009}{126}$                  \\
\hline
7  & $-\frac{1049}{2835}$               & $-\frac{13082}{945}$      & $-\frac{8059}{567}$                  \\
\hline
9  & $\frac{5585873}{5239080}$          & $-\frac{63641}{2835}$     & $-\frac{22404539}{1047816}$          \\
\hline
11 & $\frac{18407763191}{5108103000}$   & $-\frac{5128601}{155925}$ & $-\frac{149605205569}{5108103000}$   \\
\hline
\end{tabular}
\caption{Log-coefficients in odd $d$. For $d=3$ the total coefficient
$\alpha_{\text{PI}}$ agrees with \cite{Volkov:2000ih}.}
\label{tab:log}
\end{table}


\subsection{Discussion}\label{sec:discussion}

To sum up, the 1-loop graviton $S^2\times S^{d-1}$ partition function \eqref{eq:NariaiZPI} splits as
\begin{align}\label{eq:ZPIdiscuss}
    Z^\text{1-loop}_\text{PI} \left[S^2 \times S^{d-1} \right] = Z_\text{bulk} \left( \beta = \beta_N\right) Z_\text{edge} \;,
\end{align}
where
\begin{gather}\label{eq:Zbulkdiscuss}
    \log Z_\text{bulk}\left(\beta \right) =\int_0^\infty \frac{dt}{2t}\frac{1+e^{-\frac{2\pi}{\beta}t}}{1-e^{-\frac{2\pi}{\beta}t}} \chi(t) \;, \quad 
    \chi(t) \equiv \sum_{I=0}^2\sum_{l=2}^\infty \sum_{n=0}^\infty D^d_{l,I} \,\left( q^{i \omega^I_{nl}} +q^{i {\tilde\omega}^I_{nl}}\right)\;,
\end{gather}
with $q\equiv e^{- t}$, and
\begin{align}\label{eq:Zedgediscuss}
    Z_{\rm edge} &= \frac{1}{\text{vol}\left( SO(3)\times SO(d)\right)_{\text{PI}}} \left[ \det_{-1}\nolimits' \left| \frac{-\nabla_1^2-\frac{d-2}{r^2_N}}{M^2}\right|^\frac{1}{2}\det\nolimits' \left(\frac{-\nabla_0^2}{M^2}\right)^{\frac32} \right]^2 \;. 
\end{align}
Equations \eqref{eq:ZPIdiscuss}–\eqref{eq:Zedgediscuss} are the exact analogs of their $S^{d+1}$ counterparts—namely \eqref{introeq:PIsplit}, \eqref{introeq:Zbulk}, and \eqref{introeq:Zedgegravresult}.  We stress that the bulk partition function \eqref{eq:Zbulkdiscuss} can be {\it independently} defined in the Lorentzian Nariai spacetime, as elaborated in section \ref{sec:Lorentzian}.

The edge factor \eqref{eq:Zedgediscuss}, which takes the form of a path integral on $S^{d-1}$ with radius $r_N$, begs for a proper physical interpretation. Note that it receives two identical ghost-like contributions, naturally associated with the two $S^{d-1}$ horizons of Nariai. For a single copy, the simplest quadratic actions reproducing the (inverse of) determinants in \eqref{eq:Zedgediscuss} are
\begin{align}\label{eq:Nariaiedgeaction}
    S [A]  \propto \int_{S^{d-1}} \frac14 F_{ij}F^{ij} - \frac{d-2}{r^2_N} A_i A^i \;, \qquad 
    S[\chi]\propto \int_{S^{d-1}} \frac12 \partial_i \chi^A \partial^i \chi^A \;,
\end{align}
describing one tachyonic vector $A_{i}$ and three massless scalars $\chi^{A}$, $A=1,2,3$.

This marks an interesting difference from the pure dS result \eqref{introeq:Zedgegravresult}. The {\it multiplicities} of vectors and scalars coincide \eqref{introeq:Zedgegravresult}, but the scalar {\it masses} (in units of the $S^{d-1}$ radii) do not: the two tachyonic scalars with $m^{2}=-\frac{d-1}{\ell_{\text{dS}}^{2}}$ are here replaced by two massless scalars. 

When $d=3$, one might interpret \eqref{introeq:Zedgegravresult} and \eqref{eq:Nariaiedgeaction} in terms of representations of $SO(1,2)$, a common subgroup of the Lorentzian dS ($SO(1,4)$) and Nariai ($SO(1,2)\times SO(3)$) isometry groups. In that language \eqref{introeq:Zedgegravresult} contains three $\Delta=2$ (the vector plus the two tachyonic scalars) and one $\Delta=1$ (the massless scalar) discrete series representations of $SO(1,2)$ \cite{Law:2025ktz}, whereas \eqref{eq:Nariaiedgeaction} contains one $\Delta=2$ ($A_i$) and three $\Delta=1$ ($\chi^A$). For $d\geq 4$, this viewpoint is less natural since $SO(1,d-1)$ is not a common subgroup of the dS and Nariai isometry groups. If one nonetheless persists, the vector corresponds to a $\Delta = d-1$ non-unitary $SO(1,d-1)$ representation while the tachyonic and massless scalars to $\Delta =d-1$ and $\Delta =d-2$ exceptional series I representations, respectively \cite{Sun:2021thf}.

\paragraph{Shift symmetries}

In the $S^{d+1}$ case, the quadratic actions giving rise to \eqref{introeq:Zedgegravresult} are invariant under the shift symmetries \cite{Law:2025ktz} 
\begin{align}\label{eq:dSshift}
    \delta A_i = \xi^\text{KV}_i \;, \qquad \delta \phi^a = \nabla^i \xi^\text{CKV,a}_i \;, \qquad \delta \chi = c \;.
\end{align}
Here $A_i$ is the tachyonic vector, $\phi^a$ (with $a=d+1,d+2$) are the two tachyonic scalars, and $\chi$ denotes the massless scalars, all defined on an $S^{d-1}$ of radius $\ell_\text{dS}$. $\xi^\text{CKV,a}_i$ is a conformal Killing vector (CKV) on $S^{d-1}$, with $a$ labeling the two independent CKVs acting on $\phi^a$. The shift symmetries \eqref{eq:dSshift} can be seen to descend from an $SO(d)$ decomposition of $SO(d+2)$ Killing vectors on $S^{d+1}$. In contrast, \eqref{eq:Nariaiedgeaction} is invariant under 
\begin{align}
    \delta A_i = \xi^\text{KV}_i \;, \qquad \delta \chi^A = c^A \;,
\end{align}
which arise from a trivial $SO(d)$ decomposition of $SO(3)\times SO(d)$ Killing vectors on $S^2\times S^{d-1}$.

Motivated by the $S^{d+1}$ and $S^2\times S^{d-1}$ results, we conjecture that the 1-loop pure gravity path integral on any product Einstein manifold $S^p\times M_q$, with $p\geq 2,q=d+1-p\geq 3$, and compact factor $M_q$ (isometry group $H$), admit a factorization
\begin{align}\label{eq:conjecture}
    Z^\text{1-loop}_\text{PI} \left[S^p \times M_q \right] = Z_\text{bulk} \left( \beta = \beta_\text{dS}\right) Z_\text{edge} \;,\qquad \beta_\text{dS}=\frac{1}{2\pi \ell_\text{dS}} \;, \quad \ell^2_\text{dS}\equiv \frac{(d-1)(p-1)}{2\Lambda} \;. 
\end{align}
The bulk factor retains the form \eqref{eq:Zbulkdiscuss}, with $\chi(t)$ built from physical graviton QNMs on a static patch of $dS_p\times M_q$ static patch. The edge factor is the inverse path integral on $S^{p-2}\times M_q$:\footnote{The proportionality constant in \eqref{eq:Zedgeconjecture} should include the group volume $\text{vol}\left(SO(p+1)\times H \right)_\text{PI}$ and various non-local factors to ensure that $Z_\text{edge}$ is local.}
\begin{align}\label{eq:Zedgeconjecture}
    Z_{\rm edge }
    \propto 
    \begin{cases}
        \det\nolimits' \left| \frac{-\nabla_1^2-\frac{2\Lambda}{d-1}}{M^2} \right|^{\frac12}  \det\nolimits' \left| \frac{-\nabla_0^2-\frac{p}{\ell^2_\text{dS}}}{M^2} \right| \det\nolimits'  \left(\frac{-\nabla_0^2}{M^2} \right)^{\frac12}&\;, \quad p\geq 3\\
        \left[ \det\nolimits' \left| \frac{-\nabla_1^2-\frac{2\Lambda}{d-1}}{M^2}\right|^\frac{1}{2}\det\nolimits' \left(\frac{-\nabla_0^2}{M^2}\right)^{\frac32} \right]^2&\;,\quad p=2
    \end{cases}
     \;,
\end{align}
where $-\nabla^2_1$ and $-\nabla^2_0$ acts on transverse vectors and scalars on $S^{p-2}\times M_q$. The zero modes of the operators in \eqref{eq:Zedgeconjecture}--- the Killing vectors $\xi_\mu^\text{KV}=\left(\xi_a^\text{KV} ,0\right)\oplus\left(0,\xi_i^\text{KV} \right)$ of $S^{p-2}\times M_q$, the $l=1$ harmonics $Y^p_{l=1}\propto \nabla^a \xi^\text{CKV}_a$ (constant along $M_q$) and the overall constant mode---encode the non-linearly realized $SO(p+1)\times H$ symmetries. Verifying \eqref{eq:conjecture} is left to future work.

\paragraph{$S^{d-1}$ brane interpretation} 

Inspired by studies of shift-symmetric theories \cite{Goon:2011qf,Goon:2011uw,Burrage:2011bt,Clark:2005ht,Clark:2006pd,Bonifacio:2018zex,Bonifacio:2019hrj}, a possible geometric interpretation of \eqref{introeq:Zedgegravresult} was discussed in \cite{Law:2025ktz}, where the two tachyonic scalars $\phi^a$ parametrize the two coordinates transverse to an $S^{d-1}$ brane embedded in an ambient $S^{d+1}$. Expanding the world-volume action for such a brane about small  $\phi^a$ yields the quadratic action for $\phi^a$. The tachyonic vector $A_i$ and massless scalar $\chi$ can also be understood geometrically: the former describes infinitesimal diffeomorphisms on $S^{d-1}$,\footnote{In terms of the Lie derivative $M_{ij}\equiv\mathcal{L}_{A}\bar g_{ij}=\nabla_iA_j+\nabla_jA_i$, the action for $A$ is 
$S[A]\propto\int_{S^{d-1}}\bigl(\frac14 M^{ij}M_{ij}-\frac14(M^{i}{}_{i})^{2}\bigr)$.} while the latter parametrizes rotations in the normal bundle.

For $S^{2}\times S^{d-1}$ the same picture applies if we identify two of the massless scalars, say $\chi^{1},\chi^{2}$, with the two transverse coordinates to an $S^{d-1}$ brane embedded in an ambient $S^2\times S^{d-1}$. With the induced metric on the $S^{d-1}$ brane, one can show that the world-volume action in this case reduces to the massless scalar action \eqref{eq:Nariaiedgeaction} instead. For more intuition of the difference in masses, consider the spherically symmetric configurations, where the scalars are constant. In the $S^{d+1}$ case, contant translations along $\phi^a$ {\it shrink} the brane, which is why the mass $-\frac{d-1}{\ell^2_\text{dS}}$ is tachyonic. In contrast, in $S^2\times S^{d-1}$, because the $S^{d-1}$ factor has a {\it fixed} radius $r_N$, constant translations of the brane along $(\chi^{1},\chi^{2})$ leaves its size unchanged, giving massless, rather than tachyonic, scalars.

\paragraph{Distinction from $p$-form gauge theories} 

The contrast between \eqref{introeq:Zedgegravresult} and \eqref{eq:Zedgediscuss}, together with the brane picture, shows that $Z_\text{edge}$ for gravity depends not only on data intrinsic to $S^{d-1}$ but also on geometry in a {\it neighborhood} of $S^{d-1}$. 

This sharply differs from Maxwell and, more generally, $U(1)$ $p$-form gauge theories. For any static spacetime one can interpret $Z_{\text{edge}}$ in terms of ``edge modes", identified as large gauge transformations with support on the brick wall excising the co-dimension-2 horizon $\Sigma$ \cite{Ball:2024hqe,Ball:2024xhf,Ball:2024gti}. Quantizing those modes yields a thermal partition function that, in the limit where the brick wall approaches the horizon, reduces to (the inverse of) a partition function on $\Sigma$ for a compact scalar (Maxwell) or a $(p-1)$-form gauge field—objects that depend only on $\Sigma$’s intrinsic geometry.


\section*{Acknowledgments} 

It is a great pleasure to thank Luca Ciambelli, William Donnelly, Laurent Freidel, Juan Maldacena, and Gabriel Wong for stimulating conversations, and especially Dionysios Anninos, Adam Ball, and Zimo Sun for helpful discussions and comments on the draft. A.L. was supported by NSF Grant PHY-2310429, the Stanford Science Fellowship, and a Simons Investigator Award. V.L. was supported in part by the U.S. Department of Energy grant DE-SC0011941. This research was also supported in part by Perimeter Institute for Theoretical Physics. Research at Perimeter Institute is supported by the Government of Canada through the Department of Innovation, Science and Economic Development and by the Province of Ontario through the Ministry of Research, Innovation and Science.

\begin{appendix}


\section{Spherical harmonics on $S^{d-1}$}\label{sec:STSH}

\subsection{Eigenvalues, degeneracies, and induced harmonics}\label{sec:harmonics}

We collect relevant facts about scalar, vector, and tensor spherical harmonics on $S^{d-1}$ of radius $R$ in any $d\geq 3$. We refer the reader to \cite{rubin1984eigenvalues,Higuchi:1986wu} for their explicit constructions.

We start with the usual scalar spherical harmonics of total angular momentum $l\geq 0$  on $S^{d-1}$, whose eigenvalues and degeneracies are:
\begin{gather}\label{eq:scalarharmonic}
	-\nabla_0^2\, Y^{d-1}_l=\frac{l\,(l+d-2)}{R^2} Y^{d-1}_l \; , \qquad D^d_{l,0} =\frac{2l+d-2}{d-2}  \binom{l+d-3}{d-3} \;, \qquad l\geq0 \;. 
\end{gather}
The subscript 0 in $\nabla_0^2\equiv \nabla^i \nabla_i$ and the degeneracy indicates that we are dealing with scalar harmonics; analogous subscripts will be used below for vectors and tensor harmonics. For $d=3$, \eqref{eq:scalarharmonic} recovers the familiar eigenvalue $\frac{l(l+1)}{R^2}$ and degeneracy $D^3_{l,0}=2l+1$ for a spherical harmonic $Y^2_l$ on $S^2$. To fix conventions, we can define an orthonormal basis of harmonics and denote the degeneracy label (e.g. magnetic quantum numbers) collectively by $\mathbf{m}$:
\begin{align}\label{eq:Ylnormalized}
    \int_{S^{d-1}}\sqrt{g}\, d^{d-1}x \,  Y^{d-1}_{l (\mathbf{m})} Y^{d-1}_{l' (\mathbf{m}')} = \delta_{ll'} \delta_{\mathbf{m}\mathbf{m}'}\;.
\end{align}
For notational simplicity, throughout this paper we will suppress the labels $\mathbf{m}$.

Next, the transverse vector harmonics satisfy
\begin{gather}\label{eq:vecharmonic}
	-\nabla_1^2\, Y^{d-1}_{l,i}=\frac{l\,(l+d-2)-1}{R^2} \, Y^{d-1}_{l,i} \; , \qquad 
	\nabla^i Y^{d-1}_{l,i}=0 \; , \qquad l\geq 1 \;,
\end{gather}
with degeneracies
\begin{align}
		D^d_{l,1}=\frac{l\,(l+d-2)(2l+d-2)(l+d-4)!}{(d-3)!\, (l+1)!}\;.
\end{align}
When $d=3$, due to Hodge duality in 2D, $Y^{2}_{l,i}$ can be constructed from $Y^{2}_l$ through $Y^2_{l,i} \propto \epsilon_{ij} \partial^j Y^2_l$. In this case, we have $D^3_{l,1}=D^3_{l,0}=2l+1$. We also note that the first vector harmonics $Y^{d-1}_{l=1,i}$ correspond to Killing vectors on $S^{d-1}$: 
\begin{align}
    \nabla_{i}Y^{d-1}_{l=1,j}+\nabla_{j}Y^{d-1}_{l=1,i}=0 \;.
\end{align}
Finally, the symmetric transverse traceless (STT) spin-2 harmonics obey
\begin{gather}\label{STSH eq}
	-\nabla_2^2Y_{l,i j}=\frac{l\,(l+d-2)-2}{R^2} \, Y_{l,ij} \; , \qquad
	\nabla^j Y_{l,j i}=0 =Y\indices{_{l,}_i^i} \; ,\qquad l\geq 2 \;, 
\end{gather}
with degeneracies 
\begin{align}\label{eq:tensordegen}
	D^d_{l,2} = \frac{d(d-3)(l-1)(l+d-1)(d+2l-2)(d+l-4)!}{2\,(d-2)!\,(l+1)!}	\; . 
\end{align}
Note that \eqref{eq:tensordegen} vanishes when $d=3$, reflecting the fact that $Y^2_{l,ij}$ do not exist  \cite{rubin1984eigenvalues}. 

We can pick orthonormal bases for \eqref{eq:vecharmonic} and \eqref{STSH eq} normalized analogously to \eqref{eq:Ylnormalized}.

\subsubsection{Induced harmonics}

Given a scalar or vector spherical harmonic, one can construct induced tensor harmonics by taking derivatives and traceless symmetrizations. For example, given a scalar harmonic $Y^{d-1}_l$ we can construct a vector harmonic as
\begin{align}
    T^{d-1,0}_{l,i} =\nabla_{i} Y^{d-1}_l  \;, \qquad -\nabla^2 T^{d-1,0}_{l,i}=\frac{l\,(l+d-2)-(d-2)}{R^2} \, T^{d-1,0}_{l,i}\;, \qquad l\geq 1 \;.
\end{align}
This has a non-zero divergence given by
\begin{align}
    \nabla^i T^{d-1,0}_{l,i} = - \frac{l\,(l+d-2)}{R^2}Y^{d-1}_l \;. 
\end{align}
We can also construct traceless (but not transverse) tensor harmonics from the scalar and vector harmonics as
\begin{alignat}{2}
    T^{d-1,0}_{l,ij} & =\left(\nabla_{i}\nabla_{j}- \frac{g_{ij}}{d-1} \nabla^2\right)Y^{d-1}_l && \;, \qquad l\geq 2 \; ,\nn\\
    T^{d-1,1}_{l,ij} & = \nabla_{i}Y^{d-1}_{l,j}+\nabla_{j}Y^{d-1}_{l,i} &&\;, \qquad l\geq 2 \;.
\end{alignat}
These are eigenfunctions of the Laplacian
\begin{align}
    -\nabla^2 T^{d-1,0}_{l,ij}  &=\frac{l\,(l+d-2)-2(d-1)}{R^2}  \, T^{d-1,0}_{l,ij} \;, \nn\\
    -\nabla^2 T^{d-1,1}_{l,ij}  &=\frac{l\,(l+d-2)-(d+1)}{R^2}\,  T^{d-1,1}_{l,ij} \;,
\end{align}
with divergences given by
\begin{align}
    \nabla^i T^{d-1,0}_{l,ij}  &=-\frac{d-2}{d-1}\frac{(l-1) (l+d-1)}{R^2} \, T^{d-1,0}_{l,j} \;, \nn\\
    \nabla^i T^{d-1,1}_{l,ij}  &=- \frac{(l-1) (l+d-1)}{R^2} \, Y^{d-1}_{l,j} \;.
\end{align}
For our discussion later, from a scalar spherical harmonic $Y^{d-1}_l$ we can construct a transverse but not traceless tensor harmonic
\begin{align}\label{eq:transHar}
    V^{d-1,0}_{l,ij} = T^{d-1,0}_{l,ij} +\frac{d-2}{d-1}\frac{(l-1) (l+d-1)}{R^2} g_{ij} Y^{d-1}_l \;, \qquad l \geq 0\;, l \neq 1 \; . 
\end{align}
While {\it not} an eigenfunction of $-\nabla^2$, it is an eigenfunction of $D\indices{_i_j^k^l}\equiv - \delta^k_i \delta^l_j \nabla^2 - \frac{2}{R^2} g_{ij} g^{kl}$:
\begin{align}
    D\indices{_i_j^k^l} V^{d-1,0}_{l,kl} = \frac{l\,(l+d-2)-2(d-1) }{R^2}\, V^{d-1,0}_{l,ij} \;. 
\end{align}
The operator $D\indices{_i_j^k^l}$ and thus $ V^{d-1,0}_{l,kl}$ will arise when we study the spectrum of the transverse traceless (TT) Laplacian \eqref{eq:TTLaplacian} on $S^2\times S^{d-1}$ in the next section.


\subsection{Explicit vector and tensor eigenfunctions on $S^2\times S^{d-1}$}\label{sec:spec}

We present explicit eigenfunctions of the Laplace-type operators appearing in the graviton partition function on the Euclidean Nariai geometry \eqref{eq:ENariaimetric}, namely \eqref{eq:NariaiZPI}, generalizing the analysis in \cite{Volkov:2000ih} to arbitrary $d\geq 3$. We use $a,b,c$ and $i,j,k$ to denote indices on the $S^2$ and $S^{d-1}$ factors, respectively. We also use $L$ and $l$ to denote the angular momentum labels for spherical harmonics on $S^2$ and $S^{d-1}$ factors, respectively.

\subsubsection{The ghost Laplacian}\label{sec:ghosteigen}

We first study the eigenfunctions of the transverse vector Laplacian
\begin{align}
    \left(-\nabla_1^2-\frac{1}{\ell^2_N}\right) A_\mu = \lambda \, A_\mu\;, \qquad \nabla^\mu A_\mu =0 \;,
\end{align}
where we have expressed  the Ricci scalar on \eqref{eq:ENariaimetric} using \eqref{eq:RicciscalarNariai}. The eigenfunctions can be easily obtained by combining the spherical harmonics on $S^2$ and $S^{d-1}$. This has been  worked out in \cite{Grewal:2022hlo}. In the following, we present these according to $SO(d)$ UIRs.

\paragraph{Vector type} 

The first type of eigenfunctions are constructed out of the transverse vector harmonics on $S^{d-1}$
\begin{align}\label{appeq:NeigfnV}
	A_a =0 \;, \qquad A_i  = Y^2_{L}    \,Y^{d-1}_{l,i}  \; , \qquad  L \geq 0\; ,\quad   l\geq 1 \;,
\end{align}
which automatically satisfy the divergence-free condition. They have eigenvalues
\begin{align}\label{appeq:NeigvalV}
	{\tilde \lambda}^{(1)}_{L,l} = \frac{L(L+1)-1}{\ell_N^2}+\frac{l\,(l+d-2)-1}{r_N^2} 
\end{align}
and degeneracies $D^3_{L,0} D_{l,1}^{d}$. The $(L,l)=(0,1)$ modes correspond to Killing vectors on $S^{d-1}$.

\paragraph{Scalar type I and II}

We have two types of eigenfunctions constructed out of scalar spherical harmonics on $S^{d-1}$, with the explicit forms
\begin{align}\label{appeq:NeigfnS1}
	A_a=Y^2_{L,a}  \,Y_l^{d-1} \; , \qquad A_i =0\;, \qquad L \geq 1\; ,   \quad l\geq 0\;,
\end{align}
and 
\begin{align}\label{appeq:NeigfnS2}
	A_a=\partial_a Y^2_{L}  \,Y_l^{d-1} \; , \qquad A_i =  -\frac{(d-2)L(L+1)}{l\,(l+d-2)} Y^2_{L}   \, \partial_i Y_l^{d-1}
    \;, \qquad L \geq 1\; ,   \quad l\geq 1\;,
\end{align}
respectively. Since $Y^2_{L,a}$ is transverse, \eqref{appeq:NeigfnS1} is automatically divergence-free. The relative coefficient in \eqref{appeq:NeigfnS2} is chosen to ensure transversality.

Both \eqref{appeq:NeigfnS1} and \eqref{appeq:NeigfnS2} have eigenvalues
\begin{align}\label{appeq:NeigvalS1}
	{\tilde \lambda}^{(0)}_{L,l} = \frac{L(L+1)-2}{\ell_N^2}+\frac{l\,(l+d-2)}{r_N^2}
\end{align}
and degeneracies $D^3_{L,0} D^d_{l,0}$. The $(L,l)=(1,0)$ modes correspond to Killing vectors on $S^{2}$.

\subsubsection{The transverse traceless Laplacian}

We now turn to the eigenvalue problem for the operator \eqref{eq:TTLaplacian} on $S^2\times S^{d-1}$:
\begin{align}\label{eq:TTeigenprobem}
	\left( - g_{\mu\alpha }g_{\nu \rho}\nabla^2_2  -2 R_{\mu\alpha\nu\rho} \right) h^{\alpha\rho} = \lambda\, h_{\mu\nu} \;, \qquad \nabla^\mu h_{\mu\nu} =0 = g^{\mu\nu}h_{\mu\nu}\; . 
\end{align}
In explicit components, with \eqref{eq:RiemNariai} we can write the second-order equations as
\begin{align}\label{eq:2ndordereigen}
	-\nabla^2_2  h_{ab} +\frac{2}{\ell_N^2}\left( h_{ab} -g_{ab}h\indices{^c_c}\right) &= \lambda\, h_{ab}\;, \nn\\
     -\nabla^2_2  h_{ai}  &= \lambda\, h_{ai}\;, \nn\\
     -\nabla^2_2  h_{ij} +\frac{2}{r_N^2} \left(h_{ij} -g_{ij} h\indices{^k_k} \right)& = \lambda\, h_{ij} \;,
\end{align}
while the transversality and tracelessness conditions read 
\begin{align}\label{eq:transtraceeigen}
	\nabla^a h_{ab}+\nabla^i h_{ib} =0 =\nabla^a h_{aj}+\nabla^i h_{ij} \qquad \text{and} \qquad 
	h\indices{^a_a} + h\indices{^i_i} =0
\end{align}
respectively. Similar to the analysis in \cite{Volkov:2000ih}, one can solve this system with the general ansatz
\begin{align}\label{eq:TTansatz}
    h_{ab} &= \left( A_1 \, T^{2,1}_{L,ab} +A_2 \, T^{2,0}_{L,ab}+ A_3 \, g_{ab }Y^{2}_{L} \right)\, Y_l^{d-1} \nn\\
    h_{ai} &= \left( B_1 \, Y^2_{L,a} + B_2 \, T^{2,0}_{L,a} \right) Y^{d-1}_{l,i} + \left( B_3\, Y^2_{L,a}  + B_4 \, T^{2,0}_{L,a}  \right)T^{d-1,0}_{l,i}\nn\\
    h_{ij}&= Y^2_L \left( C_1\, Y^{d-1}_{l,ij}+ C_2\, T^{d-1,1}_{l,ij}+C_3 \, T^{d-1,0}_{l,ij}+C_4 \, g_{ij} Y^{d-1}_{l} \right) \;. 
\end{align}
Here all mode functions are constructed from the spherical harmonics and their induced harmonics presented in appendix \ref{sec:harmonics}, so that they are eigenfunctions of the operator $- g_{\mu\alpha }g_{\nu \rho}\nabla^2_2  -2 R_{\mu\alpha\nu\rho}$. The transversality and tracelessness conditions \eqref{eq:transtraceeigen} then impose relations among the coefficients in \eqref{eq:TTansatz}, leaving a set of linearly independent solutions. As before, we present the eigenfunctions according to whether they are $SO(d)$ tensors, vectors, or scalars.

\paragraph{Tensor type} 

The simplest eigenfunctions are constructed out of the STT harmonics
\begin{align}\label{appeq:NeigfnT}
	h_{a\mu }=0 \;, \qquad h_{ij}  = Y^2_{L}    \,Y^{d-1}_{l,ij} \; , \qquad  L \geq 0\; ,\quad     l\geq 2 \;, 
\end{align}
which satisfy the TT conditions \eqref{eq:transtraceeigen} automatically. These have eigenvalues
\begin{align}\label{appeq:NeigvalT}
	\lambda^{(2)}_{L,l} = \frac{L(L+1)}{\ell_N^2}+\frac{l\,(l+d-2)}{r_N^2} 
\end{align}
and degeneracies $D^3_{L,0} D_{l,2}^{d}$. Note that this sector is absent in $d=3$.

\paragraph{Vector type I and II}

The first type of vector eigenfunctions take the form
\begin{align}\label{appeq:NeigfnV1}
	h_{ab}=0\;, \quad h_{a i}= Y^2_{L,a}  \,Y^{d-1}_{l,i} \; , \quad h_{ij} =0\;, \quad L \geq 1\; ,   \quad l\geq 1\;,
\end{align}
which satisfy the TT conditions \eqref{eq:transtraceeigen} automatically. The second type is 
\begin{align}\label{appeq:NeigfnV2}
	h_{ab}=0\;, \quad  h_{ai}= \partial_a Y^2_{L}  \,Y^{d-1}_{l,i} \; , \quad h_{ij}=  -\frac{(d-2)L(L+1)}{(l-1) (l+d-1)}  Y^2_{L}   \, T^{d-1,1}_{l,ij}
    \;,\qquad L \geq 1\; ,   \quad l\geq 2\;,
\end{align}
where the relative coefficient between the $h_{ai}$ and $h_{ij}$ is fixed by the TT conditions \eqref{eq:transtraceeigen}.

Both \eqref{appeq:NeigfnV1} and \eqref{appeq:NeigfnV2} have eigenvalues
\begin{align}\label{appeq:NeigvalV1}
	\lambda^{(1)}_{L,l} = \frac{L(L+1)-1}{\ell_N^2}+\frac{l\,(l+d-2)-1}{r_N^2}
\end{align}
and degeneracies $D^3_{L,1} D_{l,1}^{d}=D^3_{L,0} D_{l,1}^{d}$. 

\paragraph{Scalar type I, II and III}

There are three types of eigenfunctions constructed out of scalar harmonics on $S^{d-1}$. The first two types have vanishing traces on the $S^2$ and $S^{d-1}$ factors separately:
\begin{align}\label{appeq:NeigfnTS1}
	h_{ab} &=  T^{2,1}_{L,ab} \, Y_l^{d-1}\;, \nn\\
    h_{ai } &= -\frac{(d-2)(L+2)(L-1)}{l\,(l+d-2)} Y^2_{L,a}  \, \partial_i Y_l^{d-1} 
    \;, 	\nn\\
    h_{ij}&=0 \;, \qquad\qquad\qquad\qquad\qquad\qquad\qquad\qquad L \geq 2\; ,   \quad l\geq 1\; ,
\end{align}
and
\begin{align}\label{appeq:NeigfnTS2}
	h_{ab} &= T^{2,0}_{L,ab} \, Y_l^{d-1}  \;, \nn\\
     h_{ai } &=-\frac{(d-2)(L+2)(L-1)}{2\,l\,(l+d-2)}  \partial_a Y^2_{L}  \, \partial_i Y_l^{d-1}  \; , \nn\\
    h_{ij}&=\frac{(d-2)(d-1)(L+2)(L-1)L(L+1)}{2\,l\, (l-1)(l+d-1)(l+d-2)} Y^2_{L}   \, T^{d-1,0}_{l,ij}\;, \qquad  L \geq 2\; ,   \quad l\geq 2\; . 
\end{align}

The traces on the $S^2$ and $S^{d-1}$ factors are captured by the third type of scalar eigenfunctions, which for $L\neq 1$ and $l\neq1$ can be chosen to take the form 
\begin{gather}\label{appeq:NeigfnTS3}
	h_{ab} = V^{2,0}_{L,ab} \, Y_l^{d-1}\;, \quad h_{ai } =0 \;,  \quad 
	h_{ij}= -\frac{(L+2)(L-1)}{(l-1)(l+d-1)}  Y^2_{L}   \, V^{d-1,0}_{l,ij} \;, \nn\\
	L,l \geq 0 \;, \qquad \text{except} \qquad L=1 \text{ or } l=1 \;. 
\end{gather}
For $L=1$ or $l=1$, the induced transverse harmonics \eqref{eq:transHar} do not exist. When $L=1$ and $l\neq 1$, \eqref{appeq:NeigfnTS3} is replaced by
\begin{align}
    h_{ab} &= g_{ab} Y^2_{L=1} \,  Y_l^{d-1}\;, \nn\\
    h_{ai } &= \frac{r_N^2 \partial_a Y^2_{L=1}  \, \partial_i Y_l^{d-1}}{l\,(l+d-2)}  \;,  \nn\\
	h_{ij}&=-Y^2_{L=1}   \, \left( \frac{2 \, r_N^2 \left(l\,(l+d-2) +(d-1)(d-2)\right)}{(d-2)(l-1)
   \,l\,(l+d-2) (l+d-1)} T^{d-1,0}_{l,ij} + \frac{2g_{ij} }{d-1} Y^{d-1}_l\right)\;.
\end{align}
When $L\neq 1$ and $l= 1$, we replace \eqref{appeq:NeigfnTS3} instead with
\begin{align}
    h_{ab} &=  -\left(\frac{(d-1)(L(L+1)(d-2)+2)\, l_N^2}{(d-2)L\, (L+1)(L+2)(L-1)} T^{2,0}_{L,ab} +\frac{d-1}{2} g_{ab}Y^2_{L}\right)Y_l^{d-1}\;, \nn\\
    h_{ai } &= \frac{l_N^2\partial_a Y^2_{L} \, \partial_i Y_{l=1}^{d-1}}{L(L+1)}   \;,  \nn\\
	h_{ij}&= Y^2_{L}   \, g_{ij} Y^{d-1}_{l=1} \;.
\end{align}
Note that we necessarily have $h_{ai } \neq 0$. 

All these scalar modes \eqref{appeq:NeigfnTS1}, \eqref{appeq:NeigfnTS2} and \eqref{appeq:NeigfnTS3} have the same eigenvalues
\begin{align}\label{appeq:NeigvalTS}
	\lambda^{(0)}_{L,l} = \frac{L(L+1)-2}{\ell_N^2}+\frac{l\,(l+d-2)}{r_N^2}
\end{align}
and degeneracies $D^3_{L,0} D^d_{l,0}$. 

We note that there is a single mode with $(L,l)=(0,0)$, given by
\begin{align}
	\left( h_{ab} , h_{ai }, h_{ij}\right) \propto \left(  g_{ab} \, ,\,0 \,, -\frac{2}{d-1} g_{ij} \right) \;,
\end{align}
which has a negative eigenvalue
\begin{align}\label{appeq:negativespin2}
		\lambda^{(0)}_{0,0} = -\frac{2}{\ell_N^2}=-\frac{2R}{d+1}=-\frac{4\Lambda }{d-1}\;. 
\end{align}
This corresponds to the unique TT perturbation that (at the linear order) rescales the two spheres in $S^2\times S^{d-1}$ relative to each other, while preserving the total volume. 


\end{appendix}


\bibliographystyle{utphys}
\bibliography{ref}

\end{document}